\documentclass[12pt]{JHEP3}
\usepackage{epsfig,amsmath,bm}
\usepackage{graphicx}

\newcommand{\be}{\begin{equation}}
\newcommand{\ee}{\end{equation}}
\newcommand{\bea}{\begin{eqnarray}}
\newcommand{\eea}{\end{eqnarray}}
\newcommand{\bem}{\begin{multline}}
\newcommand{\eem}{\end{multline}}
\newcommand{\beg}{\begin{gather}}
\newcommand{\eeg}{\end{gather}}

\newcommand{\as}{\alpha_s}

\def\eq#1{{Eq.~(\ref{#1})}}
\def\fig#1{{Fig.~\ref{#1}}}
\newcommand{\ben}{\begin{eqnarray*}}
\newcommand{\een}{\end{eqnarray*}}

\title{DIS on a Large Nucleus in AdS/CFT}

\author{ Javier L.\ Albacete, \ Yuri V.\ Kovchegov, \ Anastasios Taliotis
\\~~\\ Department of Physics, The Ohio State University,
Columbus, OH 43210,USA \\~~\\ E-mail addresses: \email{albacete@mps.ohio-state.edu}, 
\email{yuri@mps.ohio-state.edu}, \email{taliotis.1@osu.edu}}

\date{June 2008}

\abstract{We calculate the total cross section for deep inelastic
  scattering (DIS) on a nucleus at high energy for a strongly coupled
  ${\cal N}=4$ super Yang-Mills theory using AdS/CFT correspondence.
  In analogy to the small coupling case we argue that at high energy
  the total DIS cross section is related to the expectation value of
  the Wilson loop formed by the quark--antiquark dipole. We model the
  nucleus by a metric of a shock wave in AdS$_5$.  We then calculate
  the expectation value of the Wilson loop by finding the extrema of
  the Nambu-Goto action for an open string attached to the quark and
  antiquark lines of the loop in the background of an AdS$_5$ shock
  wave. We find three extrema of the Nambu-Goto action: the string
  coordinates at the extrema are complex-valued and are given by three
  different branches of the solution of a cubic equation.  The
  physically meaningful solutions for the total DIS cross section are
  given either by the only branch with a purely imaginary string
  coordinate in the bulk or by a superposition of the two other
  branches. For both solutions we obtain the forward scattering
  amplitude $N$ for the quark dipole--nucleus scattering. We study the
  onset of unitarity with increasing center-of-mass energy and
  transverse size of the dipole: we observe that for both solutions
  the saturation scale, while energy-dependent at lower energies, at
  very high energy becomes independent of energy/Bjorken-$x$.  The
  saturation scale depends very strongly on the atomic number of the
  nucleus as $Q_s \sim A^{1/3}$.}


\preprint{}

\begin{document}


\section{Introduction}

Over the past years there has been some significant progress made in
our understanding of deep inelastic scattering (DIS) at high
energy/small Bjorken-$x$ in the framework of the Color Glass
Condensate (CGC)
\cite{Gribov:1984tu,Mueller:1986wy,McLerran:1994vd,McLerran:1993ka,
  McLerran:1993ni,Kovchegov:1996ty,Kovchegov:1997pc,Jalilian-Marian:1997xn,
  Jalilian-Marian:1997jx, Jalilian-Marian:1997gr,
  Jalilian-Marian:1997dw, Jalilian-Marian:1998cb, Kovner:2000pt,
  Weigert:2000gi, Iancu:2000hn,Ferreiro:2001qy,Kovchegov:1999yj,
  Kovchegov:1999ua, Balitsky:1996ub, Balitsky:1997mk,
  Balitsky:1998ya,Iancu:2003xm,Weigert:2005us,Jalilian-Marian:2005jf}.
It was shown that at moderately small $x_{Bj}$ the total DIS cross
section is well-described in the quasi-classical (Glauber--Mueller
(GM)/McLerran--Venugopalan (MV))
\cite{Mueller:1989st,McLerran:1994vd,McLerran:1993ka, McLerran:1993ni}
approximation including all multiple rescatterings each taken at the
lowest (two-gluon) order. At smaller $x_{Bj}$ the energy-dependence of
the cross section comes in through the non-linear small-$x_{Bj}$
Balitsky--Kovchegov (BK) \cite{Balitsky:1996ub,
  Balitsky:1997mk,Balitsky:1998ya,Kovchegov:1999yj, Kovchegov:1999ua}
and Jalilian-Marian--Iancu--McLerran--Weigert--Leonidov--Kovner
(JIMWLK) \cite{Jalilian-Marian:1997jx, Jalilian-Marian:1997gr,
  Jalilian-Marian:1997dw, Jalilian-Marian:1998cb, Kovner:2000pt,
  Weigert:2000gi, Iancu:2000hn,Ferreiro:2001qy} quantum evolution
equations. The BK and JIMWLK evolution equations unitarize the linear
Balitsky--Fadin--Kuraev--Lipatov (BFKL) \cite{Kuraev:1977fs,Bal-Lip}
evolution equation at high energies in the large-$N_c$ limit (BK) and
beyond (JIMWLK).

The main principle of the CGC approach is the existence of a momentum
scale $Q_s$, called the saturation scale, which regulates the infrared
divergences in the total DIS cross section and in other relevant
observables for high energy collisions (for a review see
\cite{Iancu:2003xm,Weigert:2005us,Jalilian-Marian:2005jf}). The
saturation scale grows as a power of the center-of-mass energy of the
collision (a power of Bjorken $x$) as follows from BK and JIMWLK
equations, and as a power of the atomic number $A$ for DIS on the
nucleus, as can be seen in the quasi-classical GM and MV approaches.
More specifically, for small-$x_{Bj}$ evolution at the leading
logarithmic ($\as \, \ln 1/x_{Bj}$) approximation (LLA) at fixed
coupling $Q_s^2 \sim (1/x_{Bj})^{\delta \, \as} \, A^{1/3}$, where
$\as$ is the strong coupling constant and $\delta \approx 4.8$
\cite{Iancu:2002tr,Albacete:2004gw}. Thus at high energies and/or for
large nuclei the saturation scale can be large, much larger than
$\Lambda_\text{QCD}$, cutting off the non-perturbative infrared (IR)
effects. This justifies the perturbative approach and makes CGC
physics perturbative.

While the picture described above is theoretically robust and
self-consistent, the physics of CGC has not yet reached the level of
precision required to make unambiguous phenomenological predictions.
The fixed coupling LLA approaches, while giving interesting
qualitative predictions confirmed by HERA and RHIC data
\cite{Kharzeev:2000ph,Kharzeev:2001yq,Kharzeev:2002pc,Albacete:2003iq,Kharzeev:2003wz,Kharzeev:2004yx,Albacete:2007sm},
require many additional assumptions to describe the data
quantitatively. At the same time it is known that NLO corrections to
the BFKL kernel (and therefore to BK and JIMWLK kernels as well) are
large \cite{Fadin:1998py,Ciafaloni:1998gs,Balitsky:2008zz}. They come
in with a sign opposite to that of the LLA kernel, leading to cross
sections decreasing with energy and other undesirable consequences
\cite{Ross:1998xw,Kovchegov:1998ae}. It is believed that resummation
of all higher order corrections should remedy this problem. This idea
is supported by the success of resummation of collinear singularities
in the all-order BFKL kernel performed in \cite{Ciafaloni:1999yw}.
Recent calculation of the running coupling corrections to the BFKL, BK
and JIMWLK
\cite{Balitsky:2006wa,Kovchegov:2006vj,Gardi:2006rp,Kovchegov:2006wf}
and the resulting successful phenomenology both for DIS and for heavy
ion collisions at RHIC \cite{Albacete:2007sm,Albacete:2007yr} also
support the possibility that resummation of all higher order
corrections to BK and JIMWLK evolution equations would improve the
agreement with the data and resolve theoretical problems posed by NLO
BFKL calculation.

While it is not clear at all how to resum higher order corrections to
the BK and JIMWLK kernels to all orders in QCD, one turns for guidance
to other QCD-like theories, such as ${\cal N}=4$ super Yang--Mills
(SYM), where one can perform calculations in the non-perturbative
limit of large 't Hooft coupling due to the Anti-de Sitter
space/conformal field theory (AdS/CFT) correspondence
\cite{Maldacena:1997re,Gubser:1998bc,Witten:1998qj,Aharony:1999ti}.
The original work in this direction was performed by Janik and
Peschanski \cite{Janik:1999zk} (see also
\cite{Polchinski:2000uf,Polchinski:2002jw,BallonBayona:2007qr,BallonBayona:2007rs}),
who showed that at large 't Hooft coupling the scattering cross
section mediated by a single pomeron exchange, corresponding to a
single graviton exchange in the bulk of AdS$_5$, grows as the first
power of energy, $\sigma_\text{tot} \sim s$.  Comparing this result to
the LLA BFKL prediction at small coupling of $\sigma_\text{tot} \sim
s^{\frac{4 \, \as \, N_c}{\pi} \, \ln 2}$
\cite{Kuraev:1977fs,Bal-Lip}, which is the same for both QCD and
${\cal N}=4$ SYM being entirely due to gluon dynamics, one concludes
that the power of energy changes from $0$ to $1$ as 't Hooft coupling
$\lambda = 4 \pi \, \as \, N_c$ goes from being small to being large.
The result of \cite{Janik:1999zk} was further expanded in
\cite{Brower:2006ea}. It also agrees with the extrapolation of the
results of weak coupling resummations \cite{Stasto:2007uv}. The
conclusion one may draw from these results is that the cross sections
should grow fastest with energy when the coupling is large. However,
as in QCD the coupling is large at low momentum transfer $Q^2$, one
would conclude that the cross section should grow steeper with $s$ as
$Q^2$ gets lower. Such conclusion would completely contradict the
existing DIS and proton-proton collisions phenomenology
\cite{Donnachie:1998gm}, which shows that the exact opposite is true:
the growth of cross sections with energy slows down with decreasing
$Q^2$.

We believe the resolution of this puzzle may lie in the physics of
parton saturation/CGC. As one goes towards lower $Q^2$, single-pomeron
exchange approximation becomes invalid, and one has to resum multiple
pomeron exchanges. These exchanges unitarize the total cross section,
significantly reducing its growth with energy
\cite{Jalilian-Marian:2005jf}. Hence to make the strongly-coupled
dynamics consistent with phenomenology, saturation/CGC effects need to
be included in the picture as well. The first steps in this direction
were performed in \cite{Hatta:2007he,Hatta:2007cs}, where the authors
studied DIS on a static ${\cal N}=4$ SYM at strong coupling by
calculating a correlator of two $R$-currents. Here we will present a
different approach to DIS in AdS. We will argue below that the total
DIS cross section at high energy can be related to an expectation
value of the Wilson loop formed by the propagators of a quark and an
anti-quark arising in the splitting of the incoming virtual photon.
The expectation value of a Wilson loop can be found at strong coupling
using the methods outlined in \cite{Maldacena:1998im} (see also
\cite{Rey:1998bq,Rey:1998ik,Sonnenschein:1999if}). We will model the
nucleus by a shock wave metric in AdS$_5$. The expectation value of
the Wilson loop in the shock wave background is related to value of
the Nambu-Goto action at the extremal string world-sheet
\cite{Maldacena:1998im} for an open string connecting the quark and
the anti-quark lines. We therefore extremize the open string action.
Our calculation is similar to those performed in
\cite{Liu:2006ug,Liu:2006nn,Liu:2006he,Argyres:2008eg,Argyres:2006vs,Argyres:2006yz,Gubser:2007zr,Ishizeki:2008dw,Chernicoff:2006hi,Chernicoff:2008sa}.
We obtain the expectation value of the Wilson loop, which is easily
related to the forward scattering amplitude $N ({\bm r}, Y)$ of a $q
\bar q$ dipole of transverse size $\bm r$ on the nuclear target with
the total rapidity interval of $Y \sim \ln s \sim \ln 1/x_{Bj}$. The
obtained $N ({\bm r}, Y)$ is unitary and exhibits saturation
transition when the dipole transverse size becomes sufficiently large.
We extract the saturation scale from the obtained expression and
notice that at very high scattering energy $s$, corresponding to large
$Y$ or small Bjorken-$x$, the saturation scale becomes independent of
energy, contrary to the perturbative behavior outlined above.

The paper is structured as follows. In Sect. \ref{setup} we discuss
the relation of the total DIS cross section to the expectation value
of a Wilson loop. We set up the problem in AdS$_5$ space and introduce
the shock wave metric. We look for the extrema of the open string
connecting the quark and the anti-quark in the shock wave background
in Sect. \ref{sect3}. We argue in Sect. \ref{sect3} that for broad
enough shock waves the Nambu-Goto action is dominated by static string
configurations. Such approximation is justified for scattering on
nuclei with sufficiently large atomic number $A$.  We solve the static
equations of motion for the open string and find that there are three
possible complex-valued solutions for string coordinates extremizing
Nambu-Goto action in the presence of the shock wave. Similar results
are usually obtained in the quasi-classical approximation in quantum
mechanics (see Ch. 131 in \cite{LL3}). In the context of AdS/CFT
similar properties have been found in \cite{Fidkowski:2003nf}.

We study the forward scattering amplitudes $N ({\bm r}, Y)$ (or,
equivalently, the $S$-matrices $S ({\bm r}, Y)$), resulting from these
solutions in Sect. \ref{sect4}. We notice that only one branch of the
solution gives a physically meaningful unitary non-negative $N ({\bm
  r}, Y)$ for all ${\bm r}$ and $Y$. The branch has purely imaginary
string coordinates in the bulk. The corresponding $N ({\bm r}, Y)$ is
given by \eq{N1} and is plotted in \fig{ndip1}. We notice that a
slight modification of Maldacena's prescription for the calculation of
Wilson loops shown in \eq{S3} allows for a combination of two other
branches to also give a meaningful and physical amplitude $N ({\bm r},
Y)$. The resulting amplitude in shown in \eq{N3} and is plotted in
\fig{ndip2}.

The saturation scale $Q_s$ as a function of center-of-mass energy $s$
is found in Sect. \ref{sect4} as well and is shown in Figs. \ref{qs1}
and \ref{qs2} for both solutions.  For both solutions it appears that
at intermediate energies the saturation scale grows with $s$. However
the growth of $Q_s$ with $s$ stops at very high energies, reminiscent
of ``saturation of saturation'' conjectured in \cite{Kharzeev:2007zt}.
We try to interpret this result in Sect.  \ref{conc}, where we also
plot the resulting behavior of the saturation scale in the strong
coupling sector of QCD in \fig{satmap}.


\section{Setting up the Problem}
\label{setup}

In DIS at high energies, when viewed in the target rest frame, the
incoming virtual photon splits into a quark-antiquark pair, which then
scatters on the proton or nuclear target
\cite{Nikolaev:1990ja,Mueller:1989st,Kopeliovich:1993gk}. The process
is depicted in \fig{scatt}.  The scattering of a $q\bar q$ pair on the
target is eikonal at high energies
\cite{Nikolaev:1990ja,Mueller:1989st,Kopeliovich:1993gk,Kovchegov:1999yj}.
The light-cone lifetime of the $q\bar q$ pair is much longer than the
nuclear radius. Using light-cone perturbation theory
\cite{Lepage:1980fj,Brodsky:1997de} one can therefore decompose the
total scattering cross section of DIS into a convolution of a
light-cone wave function squared ($\Phi$) for a virtual photon to
decay into a $q\bar q$ pair and the imaginary part of the forward
scattering amplitude for the $q\bar q$ pair--target interaction ($N$)
\cite{Nikolaev:1990ja,Mueller:1989st,Kovchegov:1999yj}:
\begin{align}\label{sigN}
  \sigma_{tot}^{\gamma* A} (Q^2, x_{Bj}) \, = \, \int \frac{d^2 r \, d
    \alpha }{2 \pi} \, \Phi ({\bm r}, \alpha, Q^2) \ d^2 b \ N({\bm
    r}, {\bm b} , Y).
\end{align}
Here $N({\bm r}, {\bm b} , Y)$ is the imaginary part of the forward
scattering amplitude for the scattering of a quark-antiquark dipole of
transverse size $\bm r$ at center-of-mass impact parameter $\bm b$ on
a target, where the total rapidity of the scattering process is $Y =
\ln 1/x_{Bj}$ with $x_{Bj}$ the Bjorken $x$ variable.  $\Phi ({\bm r},
\alpha)$ is the light-cone wave function squared of the virtual photon
with virtuality $Q^2$ splitting into a $q\bar q$ pair of transverse
size $\bm r$ with the quark carrying a fraction $\alpha$ of the
virtual photon's longitudinal (plus) momentum. (Boldface notation
denotes two-dimensional vectors in the transverse plane.)

\FIGURE{\includegraphics[width=10.4cm]{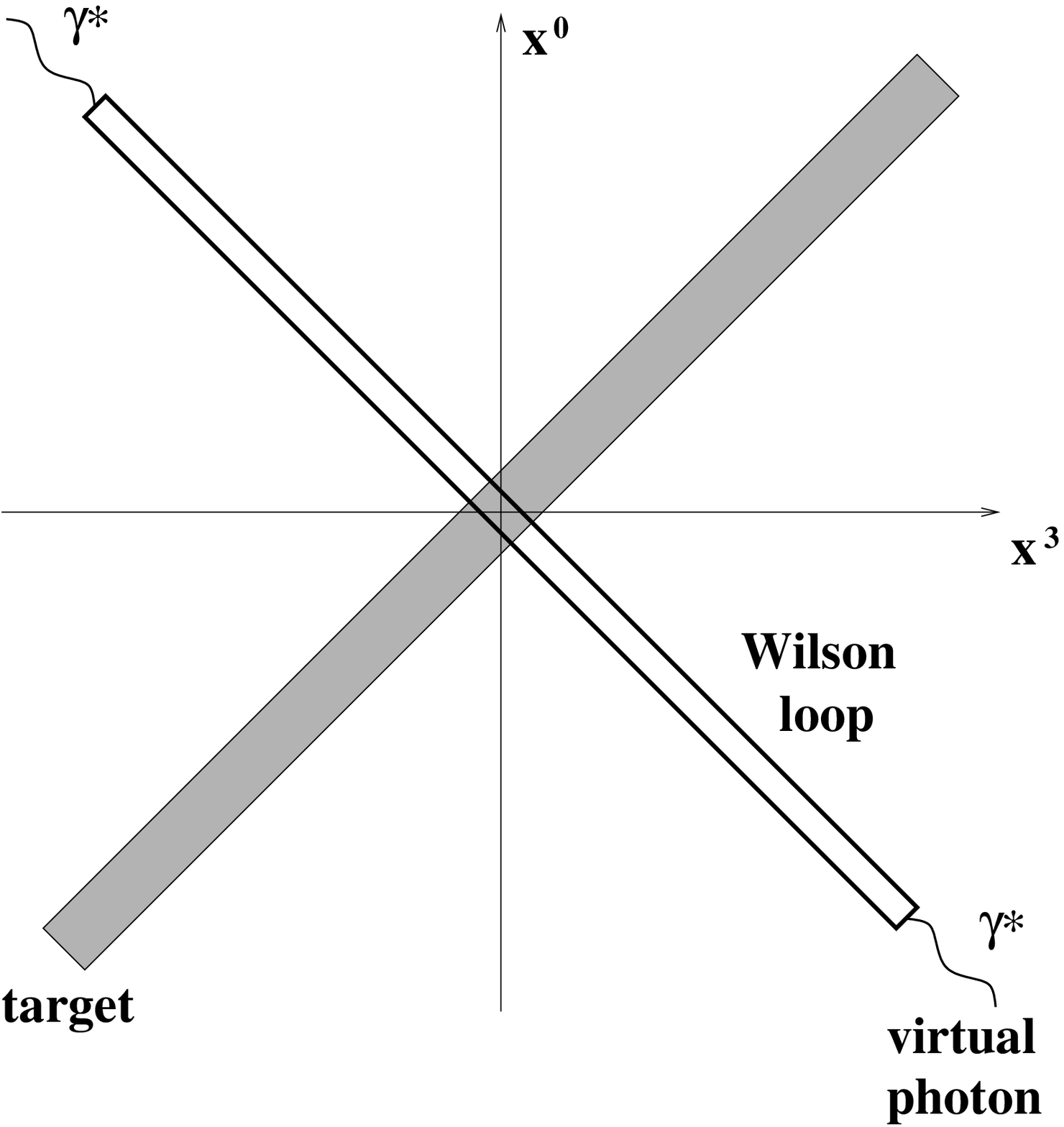}
  \caption{The space-time picture of the forward amplitude for the 
    deep inelastic scattering (DIS) at high energy. The collision axis
    is labeled $x^3$, the time is $x^0$. The virtual photon
    ($\gamma^*$) travels along one light cone and splits into a Wilson
    loop. The $q\bar q$ pair in the loop is separated only in the
    transverse plane: the separation shown along the $x^3$-direction
    is for illustration purposes only.  The Wilson loop then scatters
    on the target, traveling along the other light cone. As it is a
    forward amplitude, $q\bar q$ pair recombines back into $\gamma^*$
    after the scattering.}
  \label{scatt}
}

It is important to stress that the factorization of \eq{sigN} is
independent of the strong interaction dynamics. The wave function
squared $\Phi ({\bm r}, \alpha, Q^2)$ is purely due to
electromagnetic interactions, and is very well known
\cite{Nikolaev:1990ja,Kovchegov:1999kx,Jalilian-Marian:2005jf}. All
the strong interaction dynamics is contained in $N({\bm r}, {\bm b} ,
Y)$.

In perturbation theory, at the leading logarithmic level in $\as \,
\ln 1/x_{Bj}$, the scattering between the $q\bar q$ pair and the
target is eikonal. One can therefore relate $N({\bm r}, {\bm b} , Y)$
to the expectation value of a fundamental Wilson loop
\cite{Balitsky:1996ub,Balitsky:1998ya,Kovchegov:1999yj,Kovchegov:1999ua,Weigert:2000gi}
\begin{align}\label{N}
  N({\bm r}, {\bm b} , Y) = 1 - S({\bm r}, {\bm b} , Y)
\end{align}
with the (real part of the) $S$-matrix
\begin{align}\label{S}
  S({\bm r}, {\bm b} , Y) = \frac{1}{N_c} \, \mbox{Re} \, \bigg\langle
  W \left( {\bm b} + \frac{1}{2} {\bm r}, {\bm b} - \frac{1}{2} {\bm
      r}, Y \right) \bigg\rangle.
\end{align}
$W \left( {\bm b} + \frac{1}{2} {\bm r}, {\bm b} - \frac{1}{2} {\bm
    r}, Y \right)$ denotes a Wilson loop formed out of a quark line at
${\bm b} + \frac{1}{2} {\bm r}$ and an antiquark line at ${\bm b} -
\frac{1}{2} {\bm r}$ with the links connecting the two lines at plus
and minus infinities on the light cone. In the frame where the target
is at rest, the Wilson loop is oriented close to the light cone, with
rapidity $Y$ reflecting the ``angle'' with respect to the light cone
\cite{Balitsky:1996ub,Balitsky:1998ya}. The averaging in \eq{S},
denoted by $\langle \ldots \rangle$, is performed over all possible
wave functions of the target. 

Neglecting the gauge links at the light-cone infinities by choosing an
appropriate gauge, in QCD perturbation theory the Wilson loop can be
rewritten in terms of Wilson lines \cite{Jalilian-Marian:1997jx,
  Jalilian-Marian:1997gr, Jalilian-Marian:1997dw,
  Jalilian-Marian:1998cb, Kovner:2000pt, Weigert:2000gi,
  Iancu:2000hn,Ferreiro:2001qy,Balitsky:1996ub,Balitsky:1998ya}
\begin{align}\label{W}
  W \left( {\bm x}, {\bm y}, Y \right) \, = \, \mbox{tr} [ U ( {\bm
    x}, Y ) \, U^\dagger ({\bm y}, Y)]
\end{align}
where the Wilson line is
\begin{align}
  U ( {\bm x}, Y ) \, = \, \mbox{P} \exp \left\{ - i g \int dx_\mu \,
    A^\mu \right\}.
\end{align}
The integration contour runs from $-\infty$ to $+\infty$ close to the
light cone with the ``angle'' defined by rapidity $Y$ and $A^\mu$
being the gluon field of the target. The trace in \eq{W} is in the
fundamental representation.  In perturbative LLA, in the absence of a
target one has $A^\mu =0$, leading to $U=1$, $W=N_c$ giving $S=1$ and
$N=0$, indicating zero interaction cross section as expected. In the
opposite limit of very strong interactions one should get the black
disk limit, which corresponds to $N =1$ and $S=0$
\cite{Kovchegov:1999yj}. Hence $0 \le N, S \le 1$.

The QCD scattering problem as formulated by Eqs. (\ref{N}), (\ref{S})
and (\ref{sigN}) and shown in \fig{scatt} can be easily generalized to
$\mathcal{N}\!=\!4$ SYM theories at strong coupling by means of the
AdS/CFT prescription outlined in \cite{Maldacena:1998im,Rey:1998bq}
(for a review see \cite{Sonnenschein:1999if}). Following
\cite{Maldacena:1998im}, we represent the quark and the anti-quark in
the $\mathcal{N}\!=\!4$ SYM theory as W-bosons coming from the
breaking $U(N_c+1) \rightarrow U(N_c) \times U(1)$. Our ``quarks'' are
going to be very massive and would not recoil when interacting with
the nucleus, thus justifying the Wilson loop approximation for the
forward scattering amplitude.\footnote{While the infinite mass of the
  ``quarks'' justifies the recoilless approximation for their
  interaction with the target, the question arises of applicability of
  this model to the description of real DIS processes with light
  quarks, which indeed recoil during the interaction.  For instance,
  while at small coupling in the leading logarithmic $\as \, \ln
  1/x_{Bj}$ approximation the recoilless approximation is justified
  \cite{Balitsky:1996ub,Balitsky:1998ya,Kovchegov:1999yj,Kovchegov:1999ua},
  its validity becomes less clear at the subleading logarithmic order
  (order $\as^2 \, \ln 1/x_{Bj}$) \cite{Balitsky:2008zz}. We note that
  the recoil of the quarks at high energy only affects the impact
  factors (see e.g.  \cite{Fadin:1999qc}), and not the part of the
  interaction described by the small-$x_{Bj}$ evolution. Thus the
  calculation below, when applied to light quarks, should be
  understood as calculation of the evolution without the impact
  factors. While the impact factors are likely to be numerically
  important, we will proceed under the assumption that they will not
  affect the qualitative features of the obtained scattering
  amplitude.} The gravity dual description of the $\mathcal{N}\!=\!4$
SYM $q\bar q$ dipole corresponds to an open string in the AdS$_5$
space whose endpoints, the quark and the anti-quark, are located at
the boundary of the AdS$_5$ space.  Parameterizing the two-dimensional
world sheet of the string by the coordinates $(\tau,\sigma)$, the
location of the string in the five-dimensional world is given by
\begin{align}\label{sc}
X^{\mu}=X^{\mu}(\tau,\sigma),\quad \mu=0,\dots,4.
\end{align}
The string's Nambu-Goto action is
\begin{align}\label{SNG}
  S_{NG}=\frac{1}{2\,\pi\,\alpha'} \int d\tau \, d\sigma \,
  \sqrt{-\mbox{det} G_{\alpha\beta}}
\end{align}
with
\begin{align}
  G_{\alpha\beta} \, = \, g_{\mu\nu} (X) \,\partial_{\alpha} X^{\mu}
  \, \partial_{\beta} \, X^{\nu}, \hspace*{1cm} \alpha, \beta =
  \sigma, \tau,
\end{align}
where $g_{\mu\nu}$ will be taken below to be the metric of the AdS$_5$
space in the presence of the shock wave and $\alpha'$ is the slope
parameter.  The AdS/CFT correspondence prescription of
\cite{Maldacena:1998im} dictates that in the limit of large 't Hooft
coupling
\begin{align}\label{lambda}
\lambda=g_{\,YM}^2\,N_c
\end{align}
with $g_{\,YM}$ the Yang-Mills coupling constant, and in the
large-$N_c$ limit, the expectation value of the Wilson loop is given
by the classical Nambu-Goto action of the open string in AdS$_5$ space
\begin{align}\label{WS}
  \langle W_{\mathcal{C}} \rangle \, \sim \,
  e^{i S_{NG}},
\end{align}
where $\mathcal{C}$ represents the closed contour spanned by the quark
and the anti-quark at the string endpoints.

\FIGURE{\includegraphics[width=10.4cm]{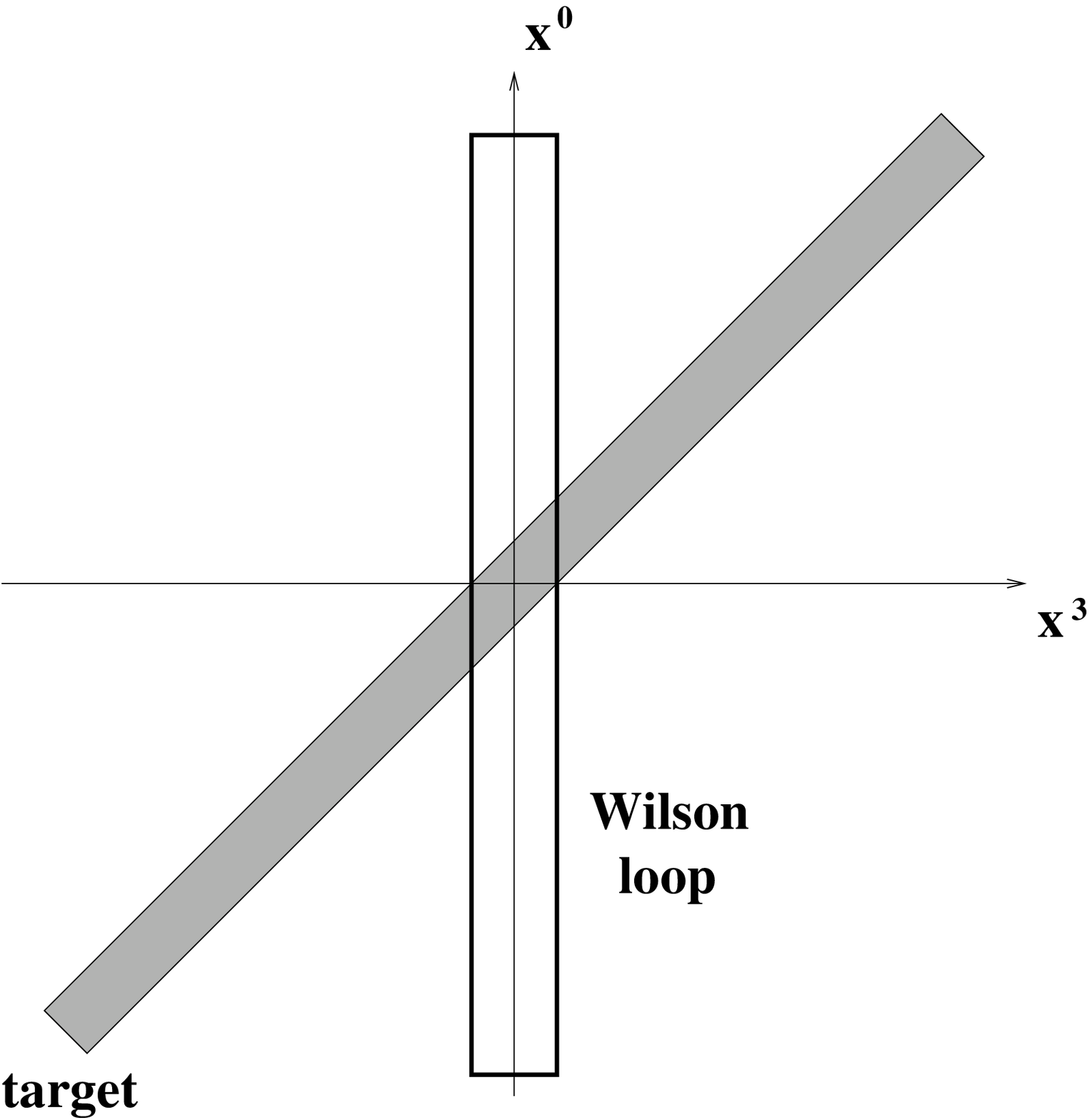}
  \caption{The same DIS process as in \protect\fig{scatt}, only viewed in the 
    rest frame of the dipole, as considered in the text. Separation of
    the quark and the anti-quark lines in $x^3$ direction was put in
    the figure only to guide the eye: in the text both the quark and
    the anti-quark are located at the same $x^3$.}
  \label{rest}
}

To complete the gravity dual description of the DIS scattering process
we still have to determine the metric of the AdS$_5$ space in the
presence of the nucleus. To that end we first introduce light-cone
coordinates 
\begin{align}\label{lc}
x^{\pm}=\frac{x^0 \pm \, x^3}{\sqrt{2}}
\end{align}
with $x^0$ the time direction.  Following Janik and Peschanski
\cite{Janik:2005zt}, we consider the ultra-relativistic nucleus as a
shock wave moving along the positive $x^3$ direction with its energy
momentum tensor given by
\begin{align}\label{t--}
 \langle T_{--} \rangle \, = \,\frac{N_c^2}{2\,\pi^2}\,\mu\,\delta(x^-),
\end{align} 
with all the other components being equal to zero. The physical
interpretation of the scale $\mu$, that has dimensions of mass cubed,
and of the factor $N_c^2$ in \eq{t--} will be discussed below.
Then, according to holographic
renormalization \cite{deHaro:2000xn}, and using Fefferman-Graham
coordinates \cite{F-G}, the gravity dual description of the CFT
energy-momentum tensor in \eq{t--} is given by the following line
element
\begin{align}\label{metriclc}
  ds^2=\frac{L^2}{z^2}\left[-2\,dx^+dx^-+\mu\,\delta(x^-)\,z^4
    \,dx^{-2}+dx_{\perp}^2+dz^2\right],
\end{align}
where $dx_{\perp}^2=(dx^1)^2+(dx^2)^2$ with $x^1$ and $x^2$ being the
coordinates transverse to the direction of motion of the nucleus, $z$
is the coordinate describing the 5th dimension and $L$ is the
curvature radius of the AdS$_5$ space. In these coordinates the
boundary of the AdS$_5$ space, and therefore the endpoints of the
string representing the dipole, is at $z=0$.

From now on we will work in the $q\bar q$ dipole's rest frame. There
the Wilson loop in \eq{S} becomes time-like, as shown in \fig{rest}.
We choose the dipole orientation to be parallel to the plane of the
shock wave. We believe such configuration is the most relevant for
studying DIS. Our choice of coordinates is such that the quark and
anti-quark are located at $x^2\!=\!x^3\!=\!z\!=\!0$ and
$x_1\!=\!\pm\,r/2$ respectively, with $r$ the dipole's transverse
size. We recall that, due to the eikonal approximation, the quark and
the anti-quark remain fixed at those spatial locations throughout the
collision process, thereby providing the appropriate Dirichlet
boundary conditions for the string motion.  Therefore, we find it
convenient to parametrize the string coordinates (\ref{sc}) in terms
of the proper time in the dipole rest frame, $t$, and $x^1 \equiv x$
in the following way:
\begin{align}
\tau\,=\,t\,,\quad \sigma=x \in\left[-\frac{r}{2},\frac{r}{2}\right],
\end{align}
\begin{align}
  X^{\mu}(\tau,\sigma)\rightarrow \left(X^0=t,\,X^1=x
    ,\,X^2=0,\,X^3=0,\,X^4=z(t,x) \right).
\end{align}
We note that the symmetries of the problem allow us to set $X^2=0$ for
all times, since nothing in the problem depends on $x^2$. Analogous
symmetry arguments indicate that the string coordinate along the
collision axis, $X^3$, is equal to zero {\it before} the collision
with the target. A priori, there is no dynamical argument allowing us
to set $X^3=0$ at all times. Indeed, the shock wave metric
(\ref{metriclc}) is strongly singular at $t=x^3$ and, therefore, it is
conceivable that after being hit by the nucleus the string develops
some kind of motion in $x^3$-direction, such as recoil or oscillation.
However, it can be proven that for the particular dynamical limit that
we are interested here, namely in the static limit that we shall
discuss in detail in Section \ref{sect3}, the dynamics of the $X_3$
component of the string coordinates is trivial, allowing us to set it
to a constant value, which we choose here to be equal to zero. In the
general time-dependent case we would not expect $X^3$ to be equal to
zero. The collision process as described in the previous paragraphs is
represented in \fig{rest}.

Before proceeding any further let us pause to interpret our choices
\eq{t--} and \eq{metriclc}, which provide the embedding of the
ultra-relativistic nucleus into the AdS$_5$ space. First of all, as it
was shown in \cite{Janik:2005zt}, the shock wave metric in
\eq{metriclc} is an exact solution of the Einstein equations in five
dimensions with negative cosmological constant $\Lambda_c=-6/L^2$,
and, therefore, provides a plausible gravity representation of an
ultra-relativistic nucleus. Secondly, let us relate the parameter
$\mu$ in \eq{t--} to some physical quantities. The energy-momentum
tensor in \eq{t--} corresponds to a nucleus made out of many massless
ultra-relativistic point particles. Thus, we envision the nucleus as a
set of $A$ nucleons, each of them consisting of $N_c^2$ {\it valence}
point-like charges moving with light-cone momentum $p^+$ along the
positive $x^+$ direction. In our model the nucleus is infinite and
homogeneous in the ($x^1,\,x^2$)-plane transverse to its direction of
motion ($x^3$). The energy-momentum tensor of an ultra-relativistic
point particle moving with momentum $p^+$ along the light cone at
transverse position ${\bm y}$ and at the other light cone coordinate
$x^- =0$ is $T_{--} (x) = p^+ \, \delta^2 ({\bm x} - {\bm y}) \,
\delta (x^-)$.  Summing over $N_c^2$ gluons in each nucleon and over
all $A$ nucleons in the nucleus and averaging over transverse
coordinates $\bm y$ we write
\begin{align}\label{tnucl}
  \langle T_{--}^{nucleus}\rangle =\frac{A \, N_c^2}{S_{\perp}} \,
  \int d^2 y \, p^+ \, \delta^2 ({\bm x} - {\bm y}) \, \delta(x^-) \,
  = \, \frac{A \, N_c^2}{S_{\perp}}\, p^+ \, \delta(x^-) \, = \,
  \frac{N_c^2}{2 \, \pi^2} \, A^{1/3} \, \Lambda^2 \, p^+ \,
  \delta(x^-)
\end{align}
where $S_{\perp} \approx A^{2/3} \, \pi \, R_N^2$ is the transverse
area of the nucleus, $R_N$ is the nucleon radius and $\Lambda^2 = 2 \,
\pi^2/ \pi R_N^2$ is a characteristic transverse momentum scale.
Comparing Eqs. (\ref{t--}) and (\ref{tnucl}) one can now identify the
scale $\mu$ in \eq{t--} in terms of the physical scales characterizing
the nucleus
\begin{align}\label{mu}
  \mu \, = \, p^{+} \, \Lambda^2 \,A^{1/3}.
\end{align}
It is worth noting that if we had not assumed the existence of $N_c^2$
{\it valence} particles per nucleon, then the perturbation to the
metric of the empty AdS$_5$ space due to the presence of the nucleus
would be suppressed by a factor of $1/N_c^2$, becoming negligible in
the strict $N_c\to\infty$ limit.

\eq{mu} will serve as our expression for $\mu$.  A more general
treatment for the energy-momentum tensor of a nucleus in the context
of high-energy scattering has been recently proposed by the authors in
\cite{Albacete:2008vs}. In particular one could argue that the
energy-momentum tensor of the nucleus at large coupling should not be
limited to only the valence quarks (gluons) contribution. The strong
gluon fields should also contribute to the energy-momentum tensor, and
their contribution is likely to be dominant at large coupling. The
energy-momentum tensor of such fields is fixed by conformal invariance
of ${\cal N} =4$ SYM theory at hand. The structure of the
energy-momentum tensor is the same as in an Abelian theory: it was
calculated in Appendix A of \cite{Albacete:2008vs}. Here we will
simply argue that such energy-momentum tensor could be approximated
by \eq{t--} with $\mu$ given by \eq{mu} times a factor of
$\sqrt{\lambda}$.  Such tensor would not qualitatively change our
discussion below, but it will modify the expression (\ref{mu}) for
$\mu$ that we will use in what follows.

Under the approximation of a homogeneous nucleus having an infinite
transverse extent the nuclear energy-momentum tensor is independent of
the transverse coordinates, as reflected in \eq{t--}.  This means that
all the relevant dynamical quantities become rotationally invariant
and independent on the impact parameter of the collision, depending
only on the dipole size, $r = |{\bm r}|$, and on the collision energy:
\begin{align}
  W \left( {\bm x}, {\bm y}, Y \right) \,\rightarrow W \left( r, Y
  \right) \,,\quad N({\bm r}, {\bm b} , Y)\rightarrow N(r, Y) \,,\quad
  S({\bm r}, {\bm b} , Y)\rightarrow S(r, Y).
\end{align}  

Let us generalize the perturbative formulation of the scattering
problem in \eq{S} to the large coupling case.  Using Eqs.  (\ref{N}), (\ref{S})
and (\ref{WS}) we write
\begin{align}\label{S2}
  S (r, Y) \, = \, \mbox{Re} \left[ \frac{\langle W
      \rangle_\mu}{\langle W \rangle_{\mu \rightarrow 0}} \right] \, =
  \, \mbox{Re} \left[ e^{i \, [S_{NG} (\mu) - S_{NG} (\mu \rightarrow
      0)]} \right]
\end{align}
and 
\begin{align}\label{N2}
  N (r, Y) \, = \, 1 - S (r, Y). 
\end{align}
In \eq{S2} $\langle W \rangle_\mu$ is the static Wilson loop pictured
in \fig{rest} and $S_{NG} (\mu)$ is the corresponding classical
Nambu-Goto action of the open string evaluated in the background of
the shock wave. $\langle W \rangle_{\mu \rightarrow 0}$ and $S_{NG}
(\mu \rightarrow 0)$ are the same quantities in the limit of $\mu
\rightarrow 0$.  Dividing by $\langle W \rangle_{\mu \rightarrow 0}$
in \eq{S2} insures that we remove the contribution to the Wilson loop
{\sl not} generated by the interaction with the shock wave
\cite{Janik:1999zk}.


\section{Static Solution}\label{sect3}

As we discussed in the previous section, the calculation of the DIS
cross-section at strong coupling under the AdS/CFT correspondence
reduces to finding the classical trajectory of the string in the
background of the shock-wave metric in \eq{metriclc}. However, finding
the time-dependent analytic solution to the Euler-Lagrange equations
associated with the Nambu-Goto action in \eq{SNG} in the background of
the metric (\ref{metriclc}) is a very difficult mathematical problem.
In this section we concentrate in a somewhat simplified problem, the
static limit, which we expect to be a good approximation for DIS on a
very large nucleus.

Following the standard procedures used in perturbative calculations
\cite{Kovchegov:1996ty,Jalilian-Marian:1997xn,Mueller:1989st} we begin
by relaxing the ultra-relativistic limit, allowing the nucleus to have
a finite longitudinal extent in the $x^-$ direction. This can be
managed by ``smearing'' the delta function in \eq{t--}:
\begin{align}\label{smear}
  \langle T_{--}\rangle=\,\frac{N_c^2}{2\,\pi^2} \,\mu\,\delta(x^-)
  \longrightarrow \,\frac{N_c^2}{2\,\pi^2}
  \,\frac{\mu}{a}\,\theta(x^-)\,\theta(a-x^-),
\end{align} 
where $\theta$ is the Heaviside step function and 
\begin{align}\label{a}
a \, \approx \, 2 \, R \, \frac{\Lambda}{p^+} \sim\frac{A^{1/3}}{p^+}
\end{align} 
is the width of the nucleus as seen from the dipole rest frame. ($R$
is the radius of the nucleus.) After the replacement in \eq{smear} the
shock wave metric in \eq{metriclc} has to be modified accordingly,
getting
\begin{align}\label{metricsm}
  ds^2=\frac{L^2}{z^2}\left[-2\,dx^+dx^-+\frac{\mu}{a}
      \,\theta(x^-)\,\theta(a-x^-)\,z^4\,dx^{-2}+dx_{\perp}^2+dz^2\right].
\end{align}
Recalling that we considered the string to be located at $x^3=0$, Eqs.
(\ref{smear}) and (\ref{metricsm}) permit the following neat
description of the temporal sequence of the collision. For $t<0$ there
is no interaction between the string and the nucleus and the string
lives in the empty AdS$_5$ space. At $t=0$ the string is hit by the
front end of the nucleus. The interaction between the nucleus and the
string continues until $t=a \, \sqrt{2}$.  At that time the nucleus
has completely passed through the string, which returns to the empty
AdS$_5$ space.  We now concentrate on times $t>0$. We will assume that
the string reaches a stationary time-independent configuration in a
short excitation time $t_e$ after being hit by the nucleus. Under that
assumption, the string will remain in such stationary configuration
during the remaining interaction time, i.e for $t_e<t<a \sqrt{2}$.
Finally, at $t>a \, \sqrt{2}$ the interaction ceases and the string is
likely to return to its original vacuum configuration after a given
relaxation time, $t_r$.  Very schematically, the contribution to the
string action due to the interaction with the nucleus,
\begin{align}
  S_{NG}^I = S_{NG} (\mu) - S_{NG} (0),
\end{align}
can be written as
\begin{align}\label{sint}
  S_{NG}^I\,=\,\left\{\int\limits_{0}^{t_e} dt\, +
    \int\limits_{t_e}^{a \, \sqrt{2}} dt +\int\limits_{a \,
      \sqrt{2}}^{a \, \sqrt{2} +t_r} dt \right\}\,\int dx
  \,\mathcal{L}
\end{align}
where $\mathcal{L}$ is the Lagrangian density. Under the additional
assumption that both the transition and relaxation times are much
smaller than the nuclear width, i.e. that
\begin{align}\label{cond}
a\,>>\,t_e\,\sim\, t_r\,,
\end{align}
one can conclude that $S_{NG}^I$ is dominated by the contribution of
the static, time-independent solution (the second term in the r.h.s of
\eq{sint}). Unfortunately, we were not able to find a reliable
parametric estimate of neither $t_e$ nor $t_s$ in terms of the
dimensionful scales of the problem, $\mu$ and $a$. 

Instead we notice that, as can be checked explicitly, the
time-dependent equations of motion for a string right after the
nucleus hit it (and right after the string leaves the nucleus) depend
only on the ratio $\mu/a$ and on the boundary conditions at $x = \pm
r/2$, i.e., on $r$. (As we will see soon, the same applies to the
static string equations of motion inside the shock wave, which do not
``know'' about the length of the shock wave.) As neither $\mu/a$ nor
$r$ depend on $A$ (see Eqs. (\ref{mu}) and (\ref{a})), we conclude
that $t_e$ and $t_r$, along with the Nambu-Goto action for the string,
are independent of $A$. Hence the second integral on the right hand
side of \eq{sint} is the only $A$-dependent term in \eq{sint} and, as
it grows with $A$, it is bound to dominate at large enough $A$.
Therefore, in the limit of very large nucleus the static approximation
will be justified.

One could also justify the static approximation in a different way by
slightly changing the physical problem. One could imagine that in DIS
the virtual photon enters the shock wave, and then fluctuates into the
$q\bar q$ pair. The $q\bar q$ pair travels through the shock wave,
and, for the forward amplitude, it would quickly recombine back into a
virtual photon before leaving the shock wave. Interaction of such a
dipole with the shock wave has to be static, as the dipole would not
``know'' about the ends of the shock wave. Integrating over the
possible splitting and recombination times would yield Eqs. (\ref{N2})
and (\ref{S2}).

In the static limit corresponding to scattering on a very large
nucleus as described above all the relevant dynamical quantities
become independent of time. We take the smeared large nucleus with the
energy-momentum tensor described in \eq{smear}. The metric
corresponding to such shock wave is given in \eq{metricsm}. To achieve
the static limit one has to remove the product of theta-functions in
\eq{metricsm}. Changing from light-cone coordinates to $(t,x^3)$
coordinates
and dropping the theta-functions in \eq{metricsm} we obtain
\begin{align}\label{metricst}
  ds^2=\frac{L^2}{z^2}\left[- \left(1-\frac{\mu}{2\,a}\,z^4\right)
    \,dt^2+\left(1+\frac{\mu}{2\,a}\,z^4\right) \, (dx^3)^2 -
    \frac{\mu}{a} \, z^4 \, dt \, dx^3 + (dx^1)^2 +
    (dx^2)^2 +dz^2\right].
\end{align} 
Let us note that the shock wave metric in \eq{metricst}, just like the
metrics in Eqs.  (\ref{metriclc}) and (\ref{metricsm}), is an exact
solution of the Einstein equations in empty AdS$_5$ space
\begin{align}
  R_{\mu\nu} + \frac{4}{L^2} \, g_{\mu\nu} = 0.
\end{align}
On the gauge theory side the metric in \eq{metricst} corresponds to a
uniform shock wave with an infinite longitudinal and transverse extent
and with the only non-zero component of the energy-momentum tensor
given by $ \langle T_{--}\rangle = \frac{N_c^2}{2\,\pi^2}
  \,\frac{\mu}{a}$.

Substituting the static metric \eq{metricst} into the Nambu-Goto
action (\ref{SNG}) and making use of the relation between the AdS$_5$
curvature radius $L$, the slope parameter $\alpha'$ and the string
coupling $\sqrt{\lambda}$
\begin{align}
\frac{L^2}{\alpha'}\,=\,\sqrt{\lambda}
\end{align}
yields
\begin{align}\label{sstat}
  S_{NG} (\mu) \,=\, \int\limits_{0}^{a \, \sqrt{2}} dt
  \int_{-r/2}^{r/2} dx \, \mathcal{L}^{static}\,,
\end{align}
where 
\begin{align}\label{lstatic}
  \mathcal{L}^{static}=\frac{\sqrt{\lambda}}{2\,\pi}\,\frac{1}{z^2}\,
  \sqrt{(1+z'^{\,2})\left(1-\frac{\mu}{2\,a}\,z^4\right)}
\end{align}
is the static Lagrangian density. $z\equiv z(x)$ is the string
coordinate along the 5th dimension of the AdS$_5$ space (henceforth we
shall drop the argument of $z$). The prime in \eq{lstatic} denotes a
derivative with respect to $x$.

We now have all the necessary ingredients to proceed to the
calculation of the classical string configuration in the static
limit. Before doing we note that 
\begin{align}
\frac{\mu}{2\,a}\approx \frac{1}{2}\,\,p^{+2}\,\Lambda^2.
\end{align}
The center of mass energy $s$ of the collision is proportional to the
product of $p^+$ and some scale characterizing the dipole. Indeed our
dipole is made out of heavy quarks with large mass $M$, so it may
appear natural to put $s \sim p^+ \, M$. However, later on we will
renormalize the Nambu-Goto action that we will obtain by subtracting
its contribution to the quark mass. Hence we will effectively remove
$M$ out of the problem. Hence the center of mass energy of the dipole
would be equal to $p^+$ times some transverse momentum scale
characterizing the dipole. We will put the latter scale to be
proportional to the scale $\Lambda$ of the nucleus, such that $s \sim
p^+ \, \Lambda$. We thus ``define''
\begin{align}
  \frac{\mu}{2\,a} \, = \, s^2.  
\end{align}

The Euler-Lagrange equation of motion associated with the Lagrangian
\eq{lstatic} reads
\begin{align}\label{EL}
  \frac{\partial}{\partial{x}}\frac{\partial{\mathcal{L}^{static}}}
  {\partial z'}-\frac{\partial{\mathcal{L}^{static}}}{\partial
    z}\,=\,0
\end{align} 
and straightforwardly leads to
\begin{align}\label{eom_stat}
z\,z''\left(1-s^2\,z^4\right)+2\, (z'^{\,2}+1) \,=\,0.
\end{align}
We need to solve \eq{eom_stat} with the boundary condition
\begin{align}\label{bc}
  z (x = \pm r/2) \, = \, 0
\end{align}
which insures that the string ends at the boundary of the AdS$_5$
space.

Using $z''=\frac{d\,z'}{d\,z}\,z'$, the first integral of
\eq{eom_stat} can be readily obtained. We get
\begin{align}\label{zp}
  z'^{\,2}\,=\,\left(\frac{z_{max}}{z}\right)^4 \,\frac{1-s^2\,z^4}
  {1-s^2\,z_{max}^4}-1\,,
\end{align}
where the constant of integration $z_{max}$ has the meaning of the
maximum extent of the string in the $z$-direction, since $z'=0$ at
$z=z_{max}$ (see \fig{string}).  $z_{max}$ is the fundamental
parameter characterizing the static solution since, as we will show
shortly, all the relevant physical quantities can be parametrized
directly in terms of $z_{max}$.

\FIGURE{\includegraphics[width=6cm]{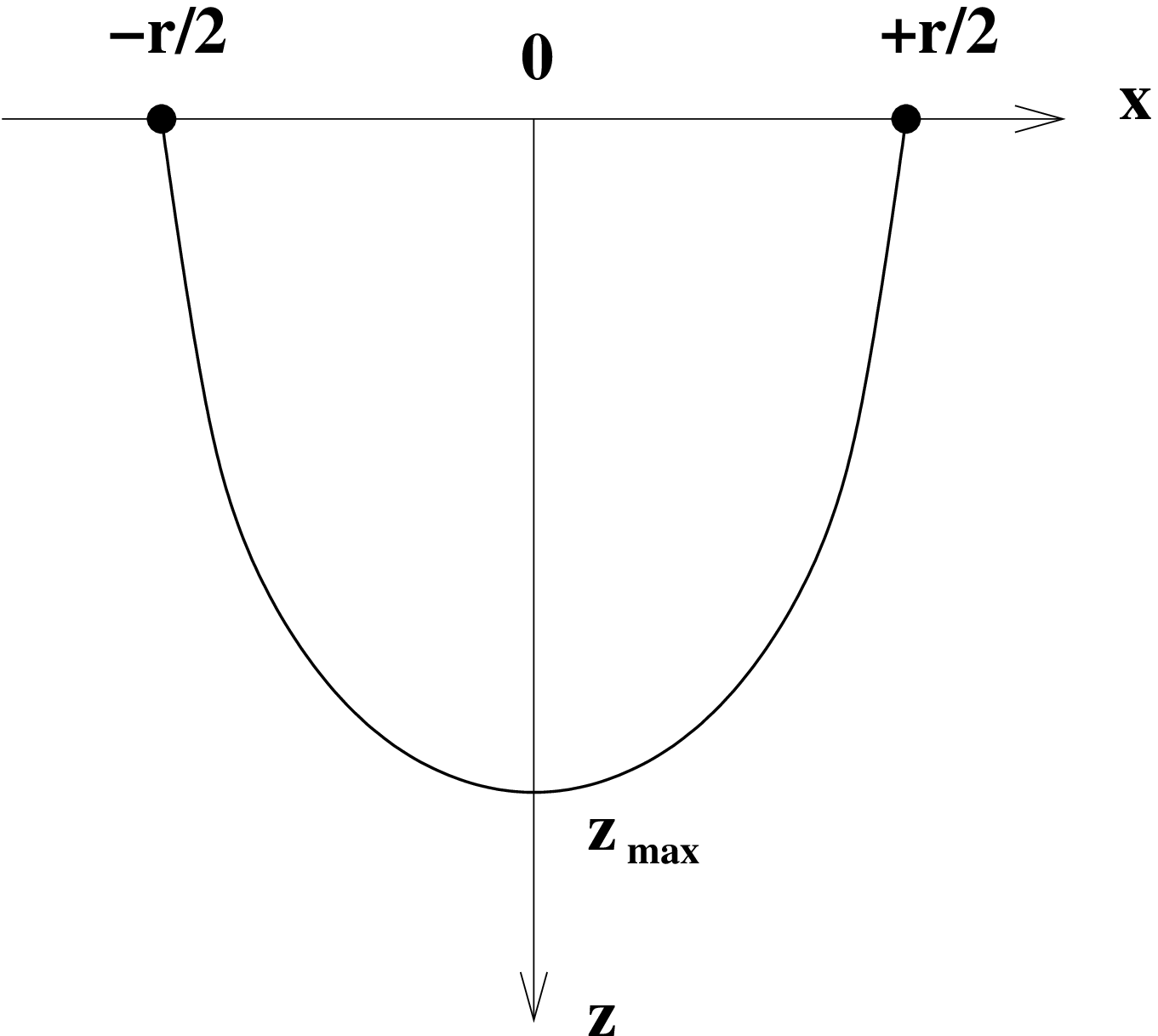}
  \caption{A sketch of the string configuration solving 
    \eq{eom_stat} and given by \eq{eom4} in the $(x,z)$-plane.}
  \label{string}
}

As $dx = dz /z'$, we write
\begin{align}\label{eom1}
  d x \, = \, \frac{dz}{\pm \sqrt{\left(\frac{z_{max}}{z}\right)^4
      \,\frac{1-s^2\,z^4} {1-s^2\,z_{max}^4}-1}}.
\end{align}
Picking the half of the string where $z' >0$ we integrate over $x$
from $-r/2$ to $x$ and, using \eq{bc} obtain
\begin{align}\label{eom2}
  x + \frac{r}{2} \, = \, \int\limits_0^z \frac{d {\tilde z}}{
    \sqrt{\left(\frac{z_{max}}{{\tilde z}}\right)^4 \,\frac{1-s^2\,
        {\tilde z}^4} {1-s^2\,z_{max}^4}-1}}.
\end{align}
Performing the integral in \eq{eom2} yields
\begin{align}\label{eom3}
  x + \frac{r}{2} \, = \, \frac{z^3}{3 \, z_{max}^3} \, z_{max} \,
  \sqrt{1-s^2\,z_{max}^4} \ F \left( \frac{1}{2}, \frac{3}{4};
    \frac{7}{4} ; \frac{z^4}{z_{max}^4} \right)
\end{align}
with $F$ the hypergeometric function. The curve in \eq{eom3} is
sketched in \fig{string}.

From symmetry considerations $z_{max}$ must lie at the midpoint
between the two string endpoints, i.e. at $x=0$. Putting $x=0$ and $z
= z_{max}$ in \eq{eom3} yields the following relation between
$z_{max}$, the collision energy $\sqrt{s}$ and the dipole size $r\,$
\begin{align}\label{zm}
  c_0 \, r \, = \, z_{max} \, \sqrt{1-s^2\,z_{max}^4}\,,
\end{align}
where we have defined a constant
\begin{align}
  c_0 \, \equiv \, \frac{\Gamma^2 \left(\frac{1}{4}\right)}{(2 \,
    \pi)^{3/2}}.
\end{align}
The case considered by Maldacena in \cite{Maldacena:1998im} for a
static Wilson loop without a shock wave can be recovered from \eq{zm}
by putting $s=0$ in it, thus eliminating the shock wave. In that limit
one gets $z_{max} = z_{max}^M \equiv c_0 \, r$.

Finally, with the help of \eq{zm}, the classical solution for the open
string in \eq{eom3} becomes
\begin{align}\label{eom4}
  x + \frac{r}{2} \, = \, \frac{z^3}{3 \, z_{max}^3} \, c_0 \, r \ F
  \left( \frac{1}{2}, \frac{3}{4}; \frac{7}{4} ; \frac{z^4}{z_{max}^4}
  \right)
\end{align}
for $-(r/2) \le x \le 0$ with a mirror image obtained by replacing $x
\rightarrow -x$ in \eq{eom4} for $0 \le x \le r/2$.


\section{Evaluating the S-matrix}\label{sect4}

We proceed by evaluating the Wilson loops in \eq{S2} using \eq{WS}
with the classical string configurations found above. As we will see
now, \eq{zm} relating the string maximum to the collision energy and
the dipole size has three different complex-valued solutions. As we
shall see below, each of the solutions leads to a different behavior
of the DIS cross-section. It turns out that the physically meaningful
solution can be obtained in two ways which are described below.


\subsection{General Evaluation}
\label{gen_eval}

We first study the solutions of \eq{zm} for the string maximum,
$z_{max}$. Introducing the dimensionless parameters
\begin{align}\label{param}
  \xi \,=\,\frac{z_{max}^2}{c_0^2\,r^2}\,,\quad \quad
  m\,=\,c_0^4\,r^4\,s^2,
\end{align}
\eq{zm} can be rewritten as the following cubic equation for $\xi$
\begin{align}\label{cubic}
  m \, \xi^3- \xi + 1 \,=\,0.
\end{align}
Obviously, for $m \neq 0$ \eq{cubic} has three different complex
roots. Solving the cubic equation (\ref{cubic}) we get
\begin{align}\label{xi}
  \xi \,=\,\frac{1}{3\,m\,\Delta}+\Delta,
\end{align}
with
\begin{align}\label{delta}
  \Delta\,=\,
  \left[-\frac{1}{2\,m}+\sqrt{\frac{1}{4\,m^2}-\frac{1}{27\,m^3}}\right]^{1/3}
  \!\exp\left[i\,\frac{2\,\pi\,n}{3}\right].
\end{align}
The index $n$ in \eq{delta} takes on values $n=0,1,2$, corresponding
to the three different Riemann sheets of the cubic root, and
characterizes the three solutions of \eq{cubic}. From here on we will
stick to the convention that the cubic root is evaluated by taking the
principal branch, while all the information distinguishing one branch
from another will be shown explicitly in terms of $n$. Also one
writes, using \eq{param},
\begin{align}\label{zmax}
  z_{max} = c_0 \, r \, \sqrt{\xi},
\end{align}
where we agree to take the principal branch of the square root. (One
can show that the secondary branch does not lead to any new physical
solutions.) The branch cut for all roots is taken along the negative
real axis.

\FIGURE{\includegraphics[width=17.5cm]{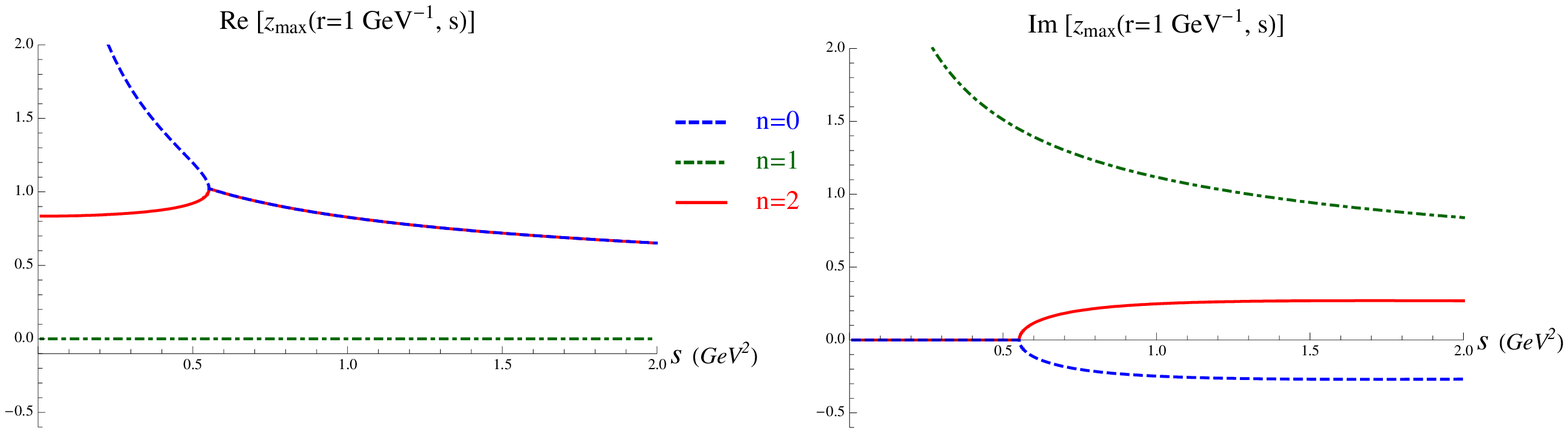}
  \caption{Real (left plot) and imaginary (right plot) parts of
    $z_{max}(r=1\, \mbox{GeV}^{-1},s)$ given by Eqs. (\ref{zmax}),
    (\ref{delta}) and (\ref{xi}) plotted as a function of the
    collision energy $s$. The figure contains all three branches of
    the solution corresponding to $n=0$ (dashed line), $n=1$
    (dot-dashed line) and $n=2$ (solid line).}
  \label{reims}
}

\FIGURE{\includegraphics[width=17.5cm]{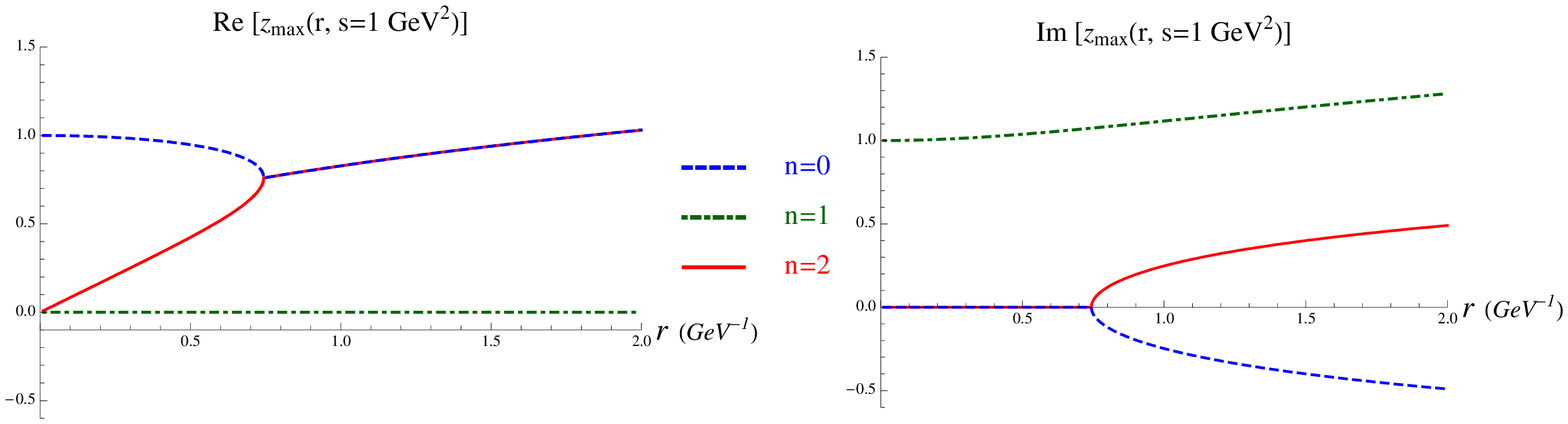}
  \caption{Real (left plot) and imaginary (right plot) parts of
    $z_{max}(r, s = 1\mbox{GeV}^2)$ given by Eqs. (\ref{zmax}),
    (\ref{delta}) and (\ref{xi}) plotted as a function of the dipole
    size $r$. Again the three branches of the solution are $n=0$
    (dashed line), $n=1$ (dot-dashed line) and $n=2$ (solid line).}
  \label{reimr}
}

The three different $z_{max}$ corresponding to the three solutions of
\eq{cubic} are shown in Figs. \ref{reims} and \ref{reimr} as functions
of energy $s$ and the dipole transverse size $r$. As shown in Figs.
\ref{reims} and \ref{reimr}, and as can be demonstrated by an
analytical calculation, one of the three solutions of \eq{cubic}, the
one corresponding to $n=1$, is purely imaginary while the other two,
corresponding to $n=0$ and $n=2$, are real for small $r$ and for small
$s$, becoming complex and conjugate to each other for larger $r$ and
$s$ once the condition of $m>4/27$ is satisfied.

Complex-valued solutions like these arise in the quasi-classical
approximations to quantum mechanics (see e.g. ch. 131 in \cite{LL3}).
They correspond to motion in the classically forbidden region. Our
physical interpretation of the complex-valued string coordinates is
that the string is classically prohibited from going through the shock
wave: such motion can only happen quasi-classically. Another way of
thinking about this result is to observe that the saddle-point
corresponding to the classical approximation does not have to lie in
the domain of real-valued string coordinates only, and can in general
be complex-valued. For similar cases in AdS/CFT framework see
\cite{Fidkowski:2003nf}.

In order to figure out which of the three branches leads to the right
physical behavior of the dipole scattering amplitude we need to know
the limiting behavior of $z_{max}$ as $m\to0$ and $m\to\infty$
respectively. A straightforward expansion of Eqs. (\ref{zmax}),
(\ref{delta}) and (\ref{xi}) yields
\begin{align}\label{asym}
  \lim_{m\to\infty}
  z_{max}=\left(\frac{c_0\,r}{s}\right)^{1/3}\left\{\begin{array}{ll}
      \exp\left[-i\frac{\pi}{6}\right],&\,\mbox{for}\,n=0\\
        i,&\,\mbox{for}\,n=1\\
      \exp\left[i\frac{\pi}{6}\right],&\,\mbox{for}\,n=2
      \end{array}\right.\,;\qquad
  \lim_{m\to0} z_{max}=\left\{\begin{array}{ll}
      1/\sqrt{s},&\,\mbox{for}\,n=0\\
      i/\sqrt{s},&\,\mbox{for}\,n=1\\
      c_0\,r,&\,\mbox{for}\,n=2
      \end{array}\right.\,.
\end{align} 
From \eq{asym} one can see right away that only the $n=2$ branch of
the solution maps onto the Maldacena result from
\cite{Maldacena:1998im} in the limit of no shock wave, i.e. as $m
\rightarrow 0$. One can also see this from Figs. \ref{reims} and
\ref{reimr}. Thus if we were to require that the string coordinates
map onto Maldacena's case for $m \rightarrow 0$, it would seem clear
that one has to take the $n=2$ branch of the solution. However, as we
will see shortly, such branch does not give a unitary $N (r,Y)$ at
large $r$: in fact $N (r,Y)$ becomes {\sl negative} at large $r$ for
the $n=2$ branch. Hence, to obtain a physically meaningful solution,
we have to either abandon the requirement of mapping the string
coordinate onto the Maldacena configuration \cite{Maldacena:1998im},
or allow for the string to ``jump'' from one branch to another. The
two possibilities are explored below.

First let us calculate the Nambu-Goto action at the extremal solutions
found above. Combining Eqs.  (\ref{sstat}) and (\ref{lstatic}) and
using $dx = dz /z'$ we write
\begin{align}\label{NG1}
  S_{NG} (\mu) \, = \, \frac{\sqrt{2 \, \lambda}\,a}{\pi}
  \int\limits_{0}^{z_{max}}
  \frac{dz}{z'}\frac{1}{z^2}\sqrt{(1+z'^2)(1-s^2z^4)}
\end{align}
where the factor of $2$ comes from adding the two halves of the
string. Using \eq{zp} in \eq{NG1} yields
\begin{align}\label{NG2}
  S_{NG} (\mu) \, = \, \frac{\sqrt{2 \, \lambda}\,a}{\pi} \, z_{max}^2
  \, \int\limits_{0}^{z_{max}} \, \frac{dz}{z^2} \, \frac{1-s^2 \,
    z^4}{\sqrt{z_{max}^4 - z^4}}.
\end{align}
The integral in \eq{NG2} has an ultraviolet divergence at $z=0$ due to
infinite quark masses. To renormalize the result one has to subtract
out the infinity \cite{Maldacena:1998im} by removing contributions of
individual quarks. The mass of a single quark in vacuum is calculated
by attaching a straight-line string to the boundary of the AdS space,
with the other end of the string going straight into the bulk.
However, similar to the case of a heavy quark potential at finite
temperature \cite{Sonnenschein:1999if}, it is not clear here whether
we should also subtract such an infinite straight string
configuration, or let the string end somewhere.

\FIGURE{\includegraphics[width=6.1cm]{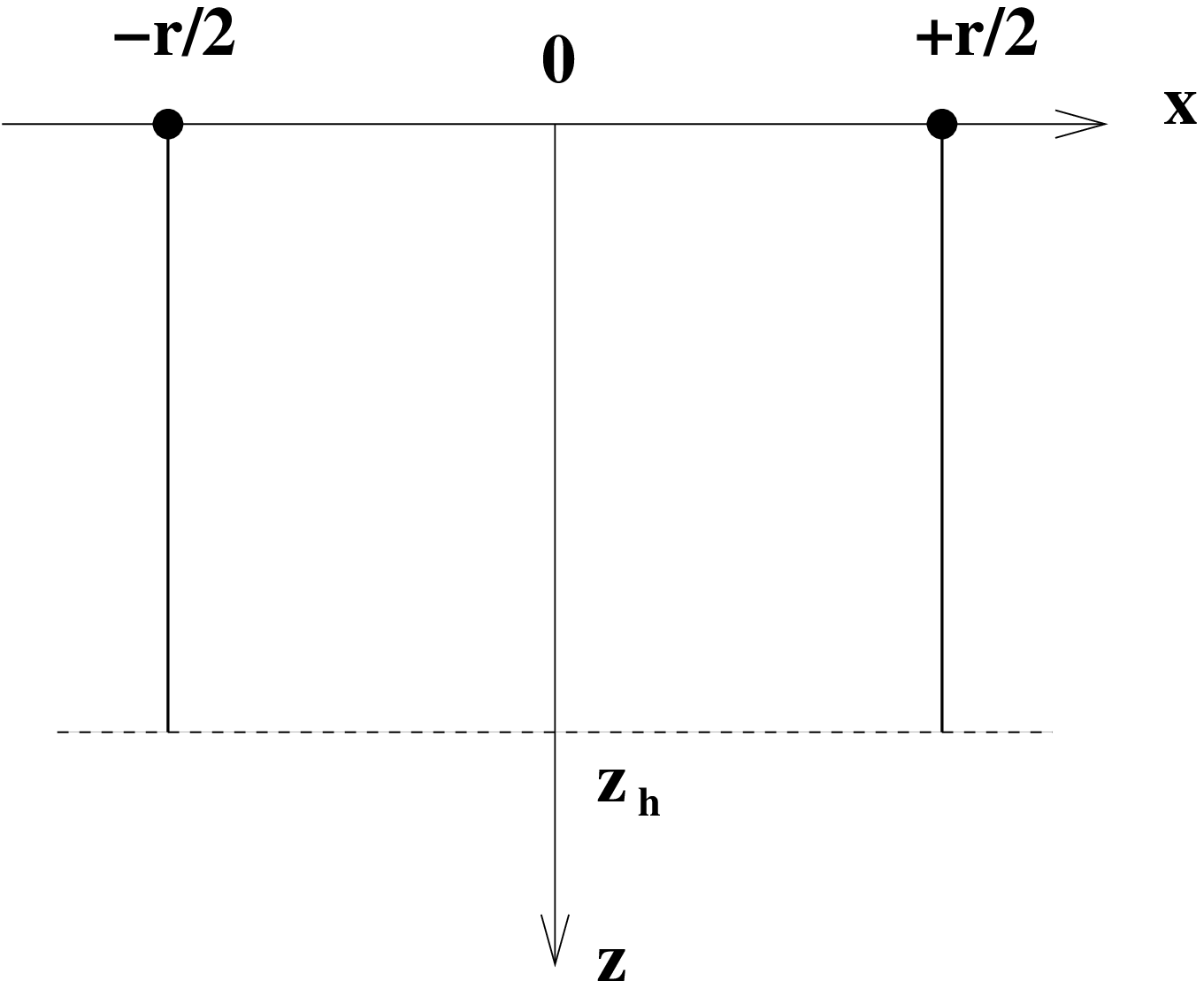}
  \caption{The string configuration which we subtract out to renormalize 
    the Nambu-Goto action drawn in the $(x,z)$-plane. The dashed line
    indicates the horizon of the shock wave.}
  \label{quarks}
}

We would like to subtract out contributions of static straight open
strings, ``hanging'' from the quark and the anti-quark into the AdS
space stretching up to $z = + \infty$. However the shock wave metric
in \eq{metricst} appears to have an analogue of a horizon. Namely, for
$z = z_h$ with the ``horizon'' defined by 
\begin{align}\label{zh}
  z_h \equiv \frac{1}{\sqrt{s}}
\end{align}
the $g_{tt}$ component of the metric becomes zero. Moreover, for $z >
z_h$, no light-like trajectories along the $z$-axis (without moving in
$x^3$ direction) are possible. Therefore the string world-sheet can not
stretch into the $z > z_h$ region. The situation is analogous to the
finite temperature calculations of heavy quark potential, where one
subtracts the contributions of strings stretching up to the black hole
horizon \cite{Rey:1998bq,Sonnenschein:1999if}. Following
\cite{Rey:1998bq,Sonnenschein:1999if} we will subtract the
straight-string configuration shown in \fig{quarks} out of the action
in \eq{NG2}. Our strings in \fig{quarks} stretch only up to the
horizon $z_h$. The action corresponding to two straight strings shown
in \fig{quarks} is obtained from \eq{NG1} by taking $z' \rightarrow
\infty$ limit in it and limiting the integration between $0 < z <
z_h$. Thus we will subtract
\begin{align}\label{subtr}
  \frac{\sqrt{2 \, \lambda} \, a}{\pi} \, \int\limits_0^{z_h}
  \frac{dz}{z^2}\, \sqrt{1-s^2\,z^4}
\end{align}
from the action of \eq{NG2}.

Subtracting \eq{subtr} out of the action in \eq{NG2} and integrating
over $z$ we obtain the renormalized action
\begin{align}\label{NGren}
  S_{NG}^{ren} (\mu) \, = \, \frac{\sqrt{\lambda}\,a}{\pi\,c_0 \,
    \sqrt{2}} \, \left[\,\frac{c_0^2\,r^2}{z_{max}^3} -
    \frac{2}{z_{max}} + \frac{2}{z_h}\,\right].
\end{align}
The action (\ref{NGren}), when substituted into Eqs. (\ref{N2}) and
(\ref{S2}) would give us the scattering amplitude $N (r,Y)$. 

Knowing the action (\ref{NGren}), and therefore the amplitude $N
(r,Y)$, we can now see which ones of the three above solutions give
physically meaningful $N (r,Y)$. Our criteria are simple: we want $N
(r,Y) \rightarrow 0$ when $r \rightarrow 0$ (color transparency) and
$N (r,Y) \rightarrow 1$ when $r \rightarrow \infty$ (the black disk
limit). In addition $N (r,Y)$ should be positive-definite for $r>0$,
$N (r,Y)$ should be less than $1$ at all $r$ (unitarity) and $N (r,Y)$
should be a monotonically increasing function of $r$ for all $r$.

A simple analysis using Eqs. (\ref{NGren}), (\ref{S2}) and (\ref{S2})
rules out the $n=2$ branch: as one can show, $N (r,Y)$ given by that
branch becomes negative at large $r$. $n=0$ branch appears to satisfy
both the color transparency condition at small $r$ and the black disk
limit at large $r$. However, the amplitude $N (r,Y)$ on that branch is
not a monotonic function and, more importantly, is not unitary: in a
certain range of energies/dipole sizes it oscillates and could become
greater than $1$ violating unitarity. Hence this branch is also
unphysical and should be discarded. This leaves us only with the $n=1$
branch, which, as we will see shortly, gives a physically meaningful
$N (r,Y)$.

The drawback of the $n=1$ branch is that it does not map onto
Maldacena's solution \cite{Maldacena:1998im} in the $\mu \rightarrow
0$ limit.  Requiring that such mapping takes place leaves us with the
$n=2$ branch, which is unphysical at large $r$. However, there is
another possibility, which we will outline below. One can take a
superposition of two branches for $z_{max}$: for small $r$ (or $s$)
one can use the $n=2$ branch, while for larger $r$ (or $s$) one can
use the $n=0$ branch.  As the $n=0$ and $n=2$ branches of $z_{max}$
intersect at $m=4/27$, as can be seen from Figs. \ref{reims} and
\ref{reimr} and from \eq{delta}, it is possible that a transition from
one branch to another happens at this point. As we will also show
below, to get a physical result from the superposition of the two
branches one has to slightly modify the rule for calculating the
contribution of the Wilson lines to the $S$-matrix (\ref{S2}). We will
consider the single $n=1$ branch solution first though.


\subsection{Single Branch}
\label{n1branch}

As one can see from \eq{asym}, and from a simple graphical analysis of
\eq{cubic}, and also from Figs.  \ref{reims} and \ref{reimr},
$z_{max}$ given by the $n=1$ branch of the solution is purely
imaginary, with a positive imaginary part. As the coordinate $x$ of
the string is real, we infer from \eq{eom4} that the complex phase of
$z$ should be the same as that of $z_{max}$. Therefore, for $n=1$
branch $z$ would be purely imaginary. Denote the positive imaginary
part of $z_{max}$ of $n=1$ branch by $\rho_{max}$ such that
\begin{align}\label{rho}
  z_{max}^{n=1} \, = \, i \, \rho_{max}, \ \ \ \rho_{max} >0. 
\end{align}
Now, as $z$ is purely imaginary, all $z$-integrals, including that in
\eq{subtr} should run along the positive imaginary axis. Therefore for
the $n=1$ branch we should replace $z_h \rightarrow i \, z_h$ in
\eq{NGren}. We rewrite \eq{NGren} as
\begin{align}\label{SNGren}
  S_{NG}^{ren} (\mu, n=1) \, = \, i \,
  \frac{\sqrt{\lambda}\,a}{\pi\,c_0 \, \sqrt{2}} \,
  \left[\,\frac{c_0^2\,r^2}{\rho_{max}^3} + \frac{2}{\rho_{max}} -
    \frac{2}{z_h}\,\right].
\end{align}
As one can explicitly check $S_{NG}^{ren} (\mu \rightarrow 0, n=1) =
0$.  Using Eqs. (\ref{S2}) and (\ref{N2}) yields
\begin{align}\label{N1}
  N (r, s) \, = \, 1 - \exp \bigg\{ -
  \frac{\sqrt{\lambda}\,a}{\pi\,c_0 \, \sqrt{2}} \,
  \left[\,\frac{c_0^2\,r^2}{\rho_{max}^3} + \frac{2}{\rho_{max}} -
    \frac{2}{z_h}\,\right] \bigg\},
\end{align}
where we replaced $Y \sim \ln s$ with $s$ in the argument of $N$.

As one can explicitly check $N (r, s)$ in \eq{N1} satisfies all
physical criteria for a meaningful solution: it goes to zero as $r
\rightarrow 0$, it approaches $1$ as $r \rightarrow \infty$ and it
varies monotonically in between, always staying positive for $r>0$ and
less than $1$. The fact that the $N (r, s)$ in \eq{N1} is a monotonic
function of $r$ for all $s$ is shown in detail in Appendix \ref{A}.

Therefore, \eq{N1} is our answer for the forward scattering amplitude
of a dipole on a nucleus given by the $n=1$ branch. We will refer to
it as Solution A. To find $\rho_{max}$ one has to use Eqs.
(\ref{rho}), (\ref{zmax}), (\ref{delta}), (\ref{xi}) and (\ref{param})
for $n=1$, which can be summarized as follows:
\begin{align}
  \rho_{max} \, = \, - i \, r \, c_0 \,
  \sqrt{\frac{1}{3\,m\,\Delta^{n=1}}+\Delta^{n=1}}
\end{align}
with
\begin{align}
  \Delta^{n=1} \, = \,
  \left[-\frac{1}{2\,m}+\sqrt{\frac{1}{4\,m^2}-\frac{1}{27\,m^3}}\right]^{1/3}
  \!\exp\left[i\,\frac{2\,\pi}{3}\right]
\end{align}
and $m\,=\,c_0^4\,r^4\,s^2$.

\FIGURE{\includegraphics[width=12cm]{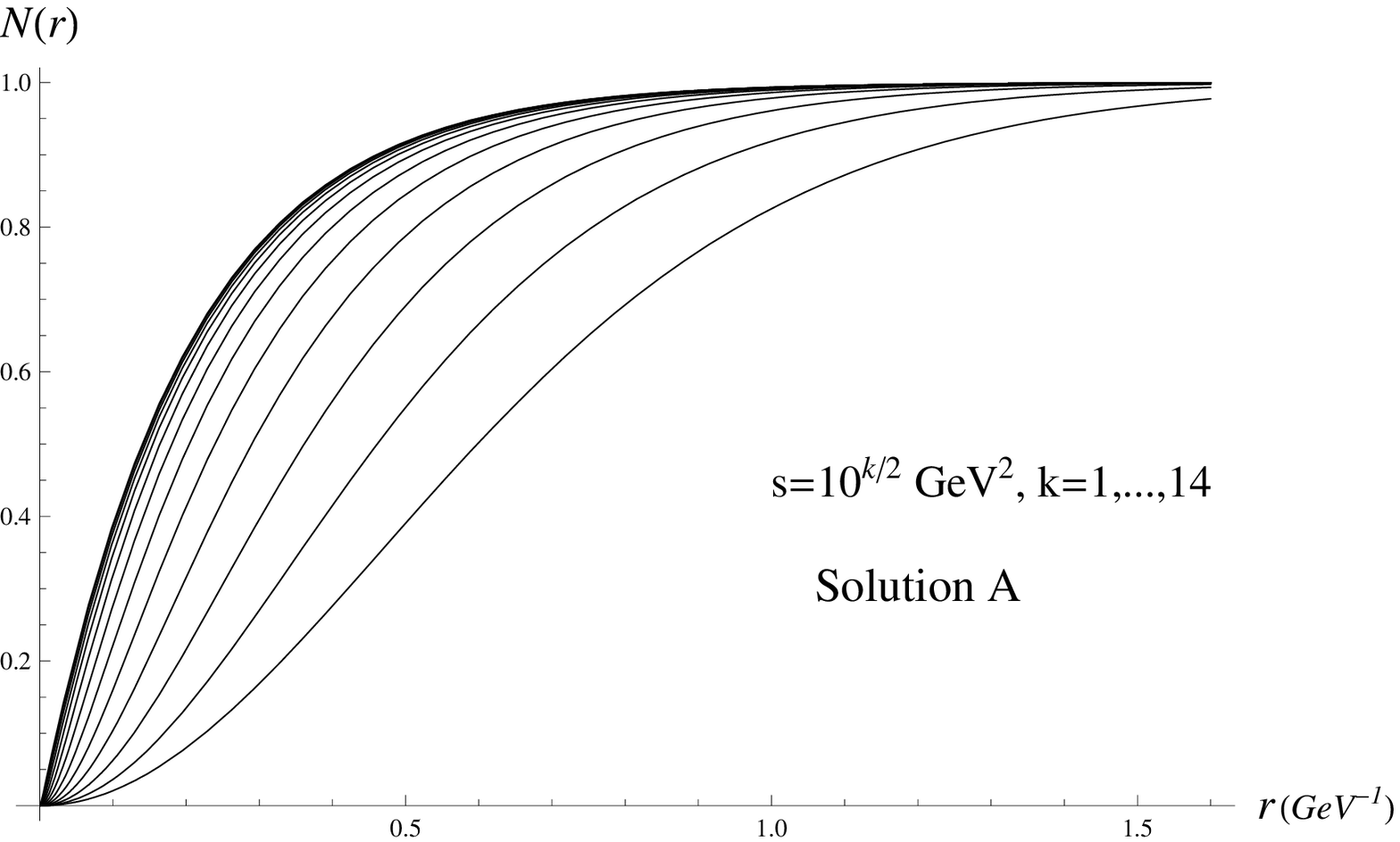}
  \caption{Dipole scattering amplitude $N(r, s)$ from \eq{N1} plotted as 
    a function of the dipole size $r$ for collision energies
    $s=10^{k/2} \,\mbox{GeV}^2,\,k=1, \ldots, 14$ (from right to left)
    with $\lambda = 20$, $A^{1/3}=5$ and $\Lambda = 1$~GeV.}
  \label{ndip1}
}

$N (r, s)$ from \eq{N1} is plotted in \fig{ndip1} as a function of $r$
for a range of center of mass energies $s$. According to \eq{a} we put
\begin{align}
  a = A^{1/3} \, \frac{\Lambda}{s}
\end{align}
with $\Lambda = 1$~GeV. We also put $\lambda = 20$ and $A^{1/3}=5$,
which is not unrealistic for DIS on a gold nucleus.  As one can see
from \fig{ndip1}, $N$ increases with increasing energy, but the growth
slows down as energy gets very high.

\FIGURE{\includegraphics[width=12cm]{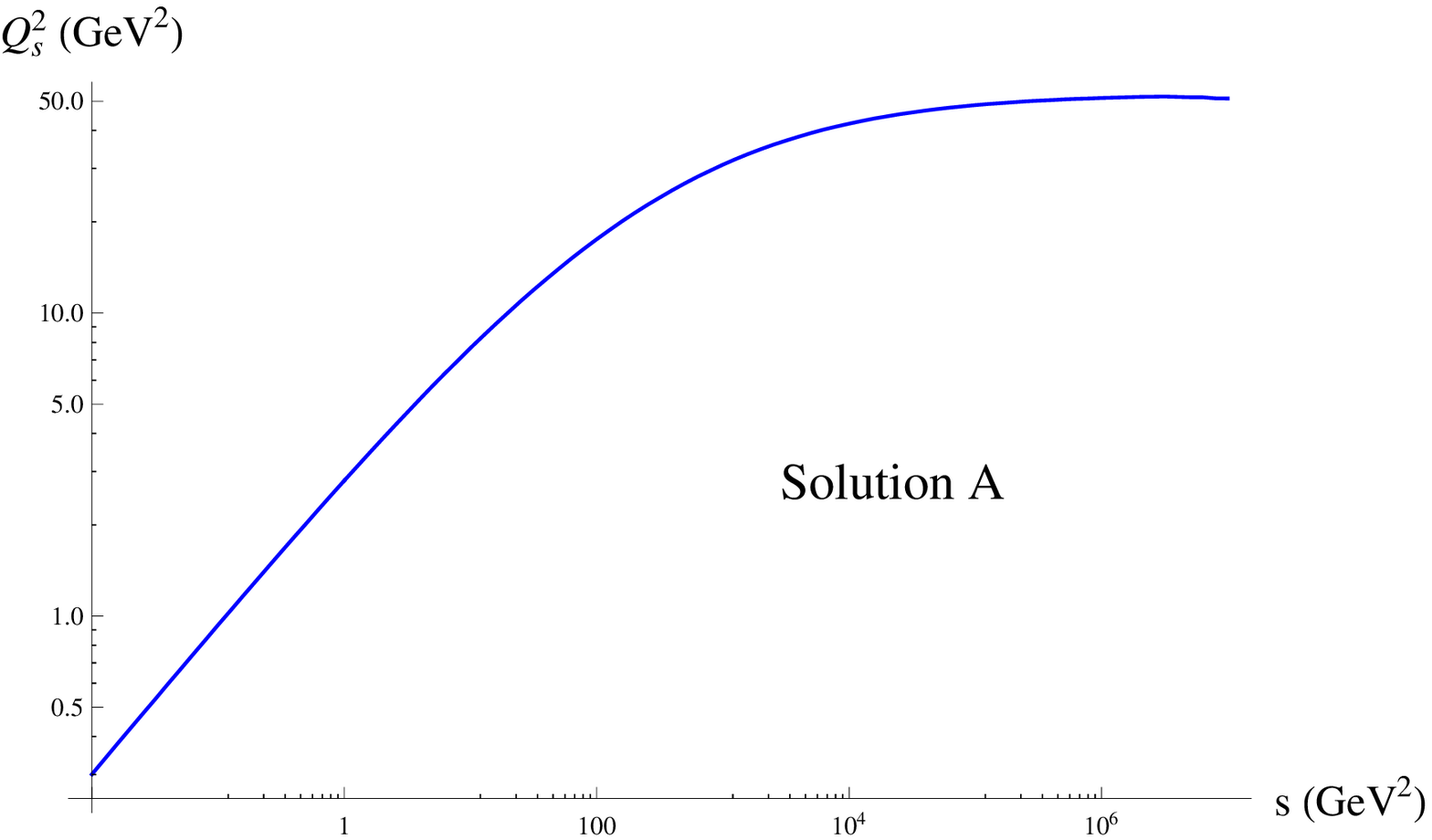}
  \caption{Saturation scale $Q_s^2$ as a function of the center-of-mass energy $s$
    corresponding to the same values of parameters as chosen for the
    solutions shown in \protect\fig{ndip1}. The saturation scale is
    obtained by imposing the condition $N(r=1/Q_s,s)=0.5$ on $N (r,s)$
    in \eq{N1}.}
  \label{qs1}
}

The saturation scale can be defined by requiring that $N (r = 1/Q_s,
s) = 0.5$. It is plotted in \fig{qs1} as a function of energy. The
saturation scale also grows with energy, but at very high energy
becomes a constant. It is interesting to note that fitting the region
of linear growth of the logarithm of the saturation scale in \fig{qs1}
with a straight line one obtains
\begin{align}\label{fit}
  \ln \frac{Q_s^2}{\Lambda^2} \, \approx \, 0.9 + 0.47 \, \ln
  \frac{s}{\Lambda^2}.
\end{align}
However the growth stops approximately at $s \approx 100$~GeV$^2$ and
$Q_s^2 \approx 20$~GeV$^2$ corresponding to $x_{Bj} \approx Q_s^2 /s
\approx 0.2$.  For higher energies/smaller $x_{Bj}$ the saturation
scale is constant with energy/$x_{Bj}$. This behavior is qualitatively
similar to the "saturation of saturation" conjectured in
\cite{Kharzeev:2007zt} using very different (small coupling) physical
arguments.

It would make more physical sense and facilitate the comparison to
other approaches if we find the dependence of the forward amplitude $N
(r,s)$ and of the saturation scale $Q_s$ not only on $s$, but also on
$x_{Bj}$. The Bjorken-$x$ variable in DIS is
\begin{align}
  x_{Bj} \, \equiv \, \frac{Q^2}{s + Q^2}
\end{align}
with $Q^2$ the virtuality of the photon. Replacing $Q \sim 1/r$
\cite{Nikolaev:1990ja} we write
\begin{align}\label{xbj1}
  x_{Bj} \, \approx \, \frac{1}{1 + s \, r^2}
\end{align}
such that for $s \, r^2 \ll 1$
\begin{align}
  x_{Bj} \, \approx \, 1 - s \, r^2
\end{align}
and for $s \, r^2 \gg 1$
\begin{align}
  x_{Bj} \, \approx \, \frac{1}{s \, r^2}.
\end{align}

To better understand our results in \eq{N1} and in Figs. \ref{ndip1}
and \ref{qs1}, let us study the asymptotics of $N (r,s)$. At lower
energies $s$ and/or for smaller dipole sizes $r$ we use the small-$m$
asymptotics (see Eqs. (\ref{param}) and (\ref{asym})) of the solution
to write
\begin{align}\label{Nsmall}
  N (r, s) \bigg|_{r^2 \, s \ll 1} \, = \, 1 - \exp \bigg\{ -
  \frac{\sqrt{\lambda} \, c_0}{2 \, \pi \, \sqrt{2}} \, r^2 \, s^{1/2}
  \, \Lambda \, A^{1/3} \bigg\}.
\end{align}
Note that in this limit $x_{Bj} \, \approx \, 1 - s r^2 \, = \, o(1)$
and the conclusions derived from \eq{Nsmall} will not apply to
small-$x_{Bj}$ physics.  The low-$s$ and low-$r$ asymptotics of the
dipole is certainly energy-dependent, as can also be seen from
\fig{ndip1}. One can use \eq{Nsmall} to extract the saturation scale:
requiring that at $r = 1/Q_s$ the power of the exponent is
approximately of order $1$ yields
\begin{align}\label{qss}
  Q_s^2 \, \sim \, \sqrt{s} \, \Lambda \, A^{1/3} \, \sqrt{\lambda}, \ 
  \ \ \ \ m \ll 1,
\end{align}
in agreement with the fit (\ref{fit}) of \fig{qs1} and the result of
\cite{Hatta:2007he}.

However, as $m \approx s^2 r^4$, the condition of $m \ll 1$ means that
$s \, r^2 \ll 1$.  
We see that \eq{Nsmall} is valid only for large $x_{Bj}$. Rewriting
\eq{Nsmall} in terms of $r$ and $x_{Bj}$ then yields
\begin{align}\label{Nsmall2}
  N (r, s) \bigg|_{r^2 \, s \ll 1} \, = \, 1 - \exp \bigg\{ -
  \frac{\sqrt{\lambda} \, c_0}{2 \, \pi \, \sqrt{2}} \, r \, \sqrt{1 -
    x_{Bj}} \ \Lambda \, A^{1/3} \bigg\}.
\end{align}
We see now that as a function of $x_{Bj}$ the saturation scale is
\begin{align}\label{qslarge}
  Q_s \, \sim \, \sqrt{1 - x_{Bj}} \ \Lambda \, A^{1/3} \,
  \sqrt{\lambda}, \ \ \ \ \ x_{Bj} \lesssim 1.
\end{align}
The expression for the saturation scale in \eq{qslarge} is valid at
large (order $1$) $x_{Bj}$ only.

At high energy and/or large dipole sizes corresponding to the
small-$x_{Bj}$ limit, \eq{N1} in the $m \gg 1$ asymptotics gives
\begin{align}\label{Nlarge}
  N (r, s) \bigg|_{r^2 \, s \gg 1} \, = \, 1 - \exp \bigg\{ -
  \frac{\sqrt{\lambda}}{\pi \, \sqrt{2}} \, r \, \Lambda \, A^{1/3}
  \bigg\}.
\end{align}
Now $x_{Bj} \approx 1/(r^2 \, s) \ll 1$ (as $s \, r^2 \gg 1$) and we
are in the small-$x_{Bj}$ regime. The saturation scale can again be
defined by requiring the power of the exponent to be of order $1$. One
then gets
\begin{align}\label{qssmall}
  Q_s \, \sim \, \sqrt{\lambda} \, \Lambda \, A^{1/3}.
\end{align}
The saturation scale is independent of energy $s$ at high energies, in
agreement with \fig{qs1}! It is also independent of $x_{Bj}$ at small
$x_{Bj}$. The saturation scale grows with the atomic number of the
nucleus $A$ as $Q_s \sim A^{1/3}$, which is a stronger growth than the
perturbative QCD estimate of $Q_s \sim A^{1/6}$. The $Q_s \sim
A^{1/3}$ scaling was also observed in \cite{Hatta:2007he}, though for
the saturation scale more similar to the one given by our \eq{qss}.
The saturation scale similar to our \eq{qssmall} was found in
\cite{Dominguez:2008vd}. Finally we note that, while in both Eqs.
(\ref{qslarge}) and (\ref{qssmall}) the saturation scale grows
proportional to $\sqrt{\lambda}$ at fixed $x_{Bj}$, the exact power of
$\lambda$ depends on the energy-momentum of the shock wave used. As we
have already mentioned above, the energy-momentum tensor of the gluon
field at strong coupling has an extra power of $\sqrt{\lambda}$ as
compared to the energy-momentum tensor of the valence gluons
(\ref{smear}) that we used above. Using the energy-momentum tensor of
the gluon field in the shock wave metric would modify the powers of
$\lambda$ in the above expressions for the saturation scales. In
particular \eq{qssmall} would then be modified to give $Q_s \sim
\lambda^{3/4}$.

Let us stress that, while the solution presented in this Subsection
and given by \eq{N1} is one of the two possible physical solutions for
$N (r,s)$, we believe that this solution is actually the correct one.
As we will see below, the amplitude $N (r,s)$ given by the $n=1$
branch does not have discontinuous $r$-derivative, unlike $N (r,s)$
given by the superposition of the other two branches. However, we do
not have a solid physical argument to select the $n=1$ solution over
the other one.


\subsection{Superposition of Two Branches}
\label{n20branch}

It is important to note that to derive our above result given by
\eq{N1} we had to abandon the condition that the string coordinates
map smoothly back onto Maldacena's vacuum solution
\cite{Maldacena:1998im} when $\mu \rightarrow 0$. Indeed the only
branch of our solution satisfying such a condition, the $n=2$ branch,
would give a physically meaningless negative $N(r,s)$ at large $r$.
However, if one insists on smooth matching with the vacuum string
configuration of \cite{Maldacena:1998im}, one can construct a
physically meaningful amplitude $N(r,s)$ using a superposition of two
branches: one can use the $n=2$ branch of the solution given by Eqs.
(\ref{xi}), (\ref{param}) and (\ref{delta}) for $m < 4/27$ and the
$n=0$ branch for $m > 4/27$. One can see from Eqs. (\ref{xi}),
(\ref{param}) and (\ref{delta}) that the values of $z_{max}$ given by
those two branches are equal at $m=4/27$, which insures smooth
matching of the values of the Nambu-Goto action between the two
solutions. (The fact that the two branches intersect can also be seen
in Figs. \ref{reims} and \ref{reimr}.)

There is a subtlety here though. If one uses the Nambu-Goto action
from \eq{NGren} for the superposition of $n=2$ and $n=0$ branches
along with Eqs. (\ref{S2}) and (\ref{N2}) to find $N (r,s)$, the
resulting amplitude $N (r,s)$ would still not be physical. In fact, in
a range of values for $s$ and $r$ it still has oscillations, which
make $N (r,s)$ go above $1$ and violate unitarity. These are the same
oscillations that we have mentioned for the $n=0$ branch above. The
origin of such oscillations is in the fact that $z_{max}$ for the
$n=0$ branch has a non-zero real part (see Figs. \ref{reims} and
\ref{reimr}), leading to a non-zero real part for the Nambu-Goto
action, which, when substituted into \eq{S2} gives cosine-like
oscillations. However, such oscillations can be removed by modifying
the prescription for the calculation of the $S$-matrix in \eq{S2}. One
can replace \eq{S2} with
\begin{align}\label{S3}
  S (r, Y) \, = \, e^{- \text{Im} [S_{NG} (\mu) - S_{NG} (\mu
    \rightarrow 0)]}.
\end{align}
Indeed by explicitly keeping the imaginary part of the Nambu-Goto
action in the exponent, as shown in \eq{S3}, we would get rid of the
real part and, therefore, of the oscillations described above. Also,
the new prescription in \eq{S3} would not change our above conclusions
in Sect. \ref{n1branch} as the action of the $n=1$ branch is purely
imaginary.  However, we do not have a good physical justification of
\eq{S3} in general. In the quasi-classical approximation to quantum
mechanics one usually takes only the imaginary part of the action to
estimate scattering amplitudes along the complex quasi-classical
trajectories, as could seen from Eq. (131.14) in \cite{LL3}. Hence our
prescription (\ref{S3}) here is motivated by quantum mechanics and by
the fact that it gives physically meaningful results here. It may be
that for calculating the $S$-matrices, the prescription of
\cite{Maldacena:1998im}, which was originally proposed for heavy quark
potentials, has to be modified to that of \eq{S3}.

Substituting \eq{NGren} into \eq{S3}, and using \eq{N2} yields
\begin{align}\label{N3}
  N (r, s) \, = \, 1 - \exp \left\{ -
    \frac{\sqrt{\lambda}\,a}{\pi\,c_0 \, \sqrt{2}} \, \text{Im}
    \left[\,\frac{c_0^2\,r^2}{z_{max}^3} - \frac{2}{z_{max}} + \frac{2
        \, | z_{max}|}{z_h \, z_{max}} \,\right] \right\},
\end{align}
where we have modified the $2/z_h$ term in the square brackets to
reflect the fact that the complex phase of $z_h$ (and of all other
$z$'s) should be the same as the complex phase of $z_{max}$, as
follows from \eq{eom4} and as we have noticed and used above in Sect.
\ref{n1branch}. $z_h$ is still given by \eq{zh} above.

\eq{N3} is the main result of this section. We will refer to it as
Solution B. When using it one should take $z_{max}$ from the $n=2$
branch of Eqs. (\ref{param}), (\ref{xi}) and (\ref{delta}) for $m <
4/27$ and from the $n=0$ branch of the same equations for $m > 4/27$.
One can also notice that $z_{max}$ from both $n=2$ and $n=0$ branches
is purely real for $m<4/27$ (see Figs.  \ref{reims} and \ref{reimr}).
Hence in practice \eq{N3} can be used with the $n=0$ branch of
$z_{max}$ for all $m$. This would eliminate the need for quantum
corrections to justify the transition between the branches. However,
the $n=0$ branch does not map onto Maldacena's vacuum solution
\cite{Maldacena:1998im}: hence by keeping $n=0$ branch only we would
also lose some of the justification for searching for solutions beyond
the $n=1$ branch described above.

Let us also note that, since $z_{max}$ and, therefore, the string
coordinates for this solution are real for $m<4/27$ and become complex
for $m > 4/27$, one can interpret $m=4/27$ as the point when, in a
purely classical sense, the string would break due to high energy of
the shock wave. This interpretation is similar to the finite
temperature case \cite{Rey:1998bq,Liu:2006nn}.

The amplitude in \eq{N3} is plotted in \fig{ndip2} for the same set of
parameters as the amplitude in \fig{ndip1}.
\FIGURE{\includegraphics[width=12cm]{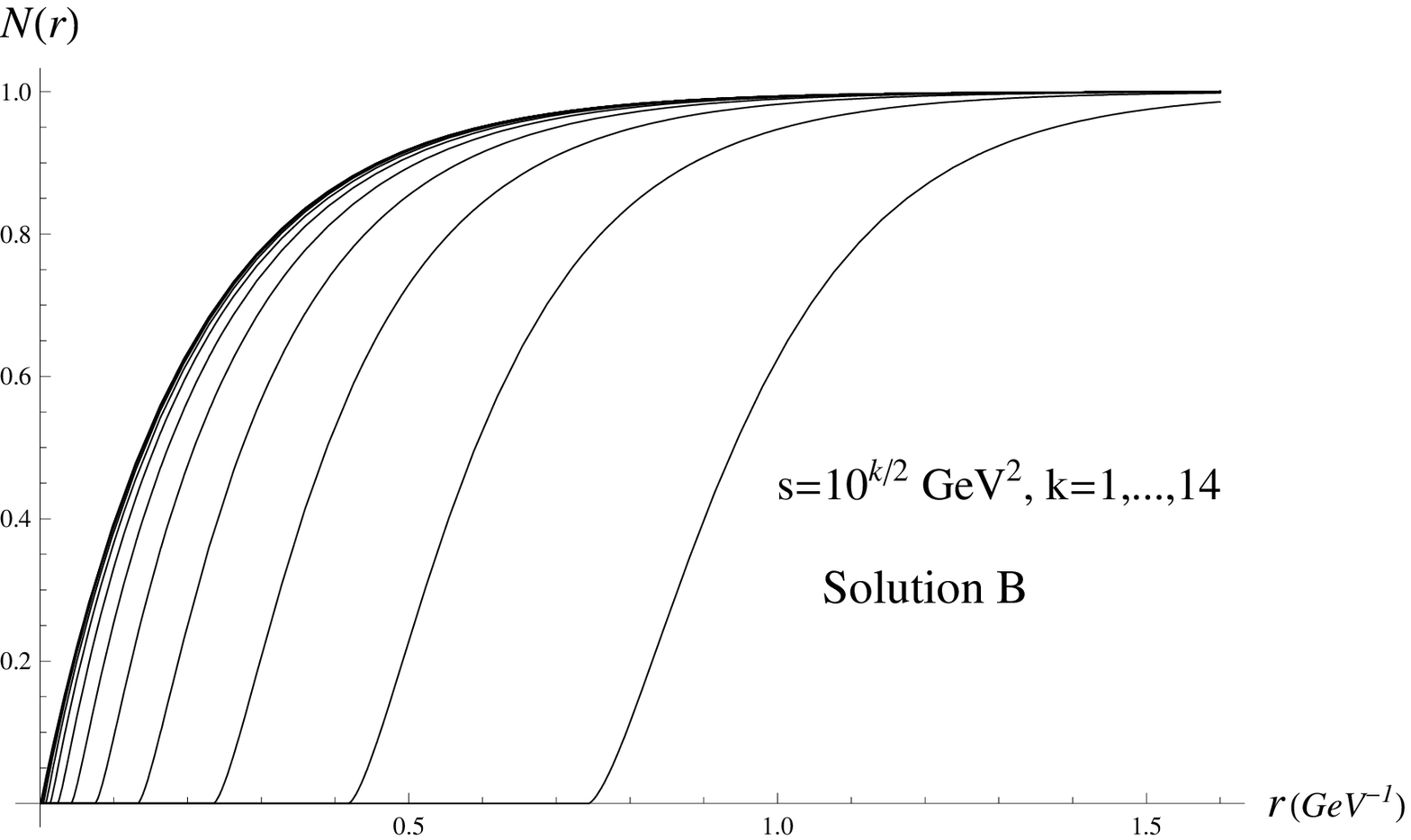}
  \caption{Dipole scattering amplitude $N(r, s)$ from \eq{N3} plotted as 
    a function of the dipole size $r$ for collision energies
    $s=10^{k/2} \,\mbox{GeV}^2,\,k=1, \ldots, 14$ (from right to left)
    with $\lambda = 20$, $A^{1/3}=5$ and $\Lambda = 1$~GeV.}
  \label{ndip2}
}
The striking feature of the amplitude in \fig{ndip2} is that $N
(r,s)=0$ for a range of non-zero values of $r$. This means that
non-zero transverse size $q \bar q$ dipoles below certain critical
transverse size would not interact with the shock wave. While such
behavior appears somewhat unphysical, one may interpret it by arguing
that at large coupling most partons are located at small-$x_{Bj}$. As
small $r$ corresponds to large $x_{Bj}$ (see \eq{xbj1}), the
small-size dipole does not resolve any partons, and thus does not
interact. The exponential suppression of high-$Q^2$ (small-$r$)
partons has been noticed before in
\cite{Polchinski:2002jw,Polchinski:2000uf,Hatta:2007cs,Hatta:2007he,Dominguez:2008vd}.
Such exponential suppression possibly translates into $0$ in our case.

\eq{N3} can be used to find the saturation scale by requiring again
that $N (r = 1/Q_s, s) = 0.5$. It is plotted in \fig{qs2} as a
function of energy. Similar to the $n=1$ branch in \fig{qs1}, the
saturation scale also grows with energy, but at very high energy
becomes a constant.
\FIGURE{\includegraphics[width=12cm]{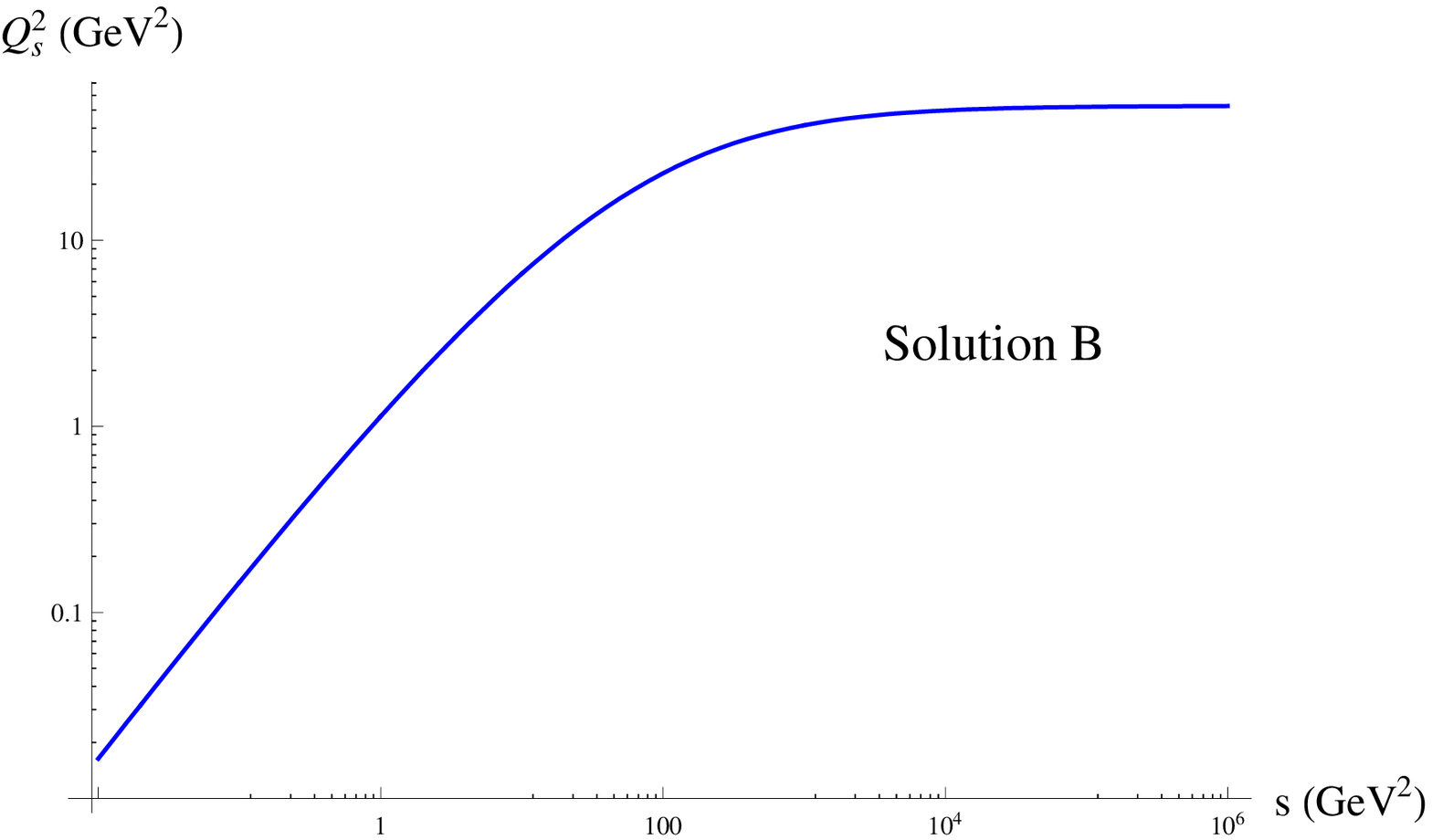}
  \caption{Saturation scale $Q_s^2$ as a function of the center-of-mass 
    energy $s$ for the amplitude plotted in \protect\fig{ndip2} and
    corresponding to the same values of parameters as chosen for the
    solutions shown in \protect\fig{ndip2}. The saturation scale is
    obtained by imposing the condition $N(r=1/Q_s,s)=0.5$ on $N (r,s)$
    in \eq{N3}.}
  \label{qs2}
}
Fitting the growing part of the curve in \fig{qs2} with a straight
line yields
\begin{align}\label{fit2}
  \ln \frac{Q_s^2}{\Lambda^2} \, \approx \, 0.25 + 0.95 \, \ln
  \frac{s}{\Lambda^2}.
\end{align}
Again the growth of the saturation scale stops at a rather large
values of $x_{Bj}$.

Let us study the low and high energy asymptotics of \eq{N3}. $N (r,s)$
in \eq{N3} and in \fig{ndip2} becomes non-zero at $m = 4/27$
corresponding to
\begin{align}
  r \, = \, r^* \, \equiv \, \left( \frac{4}{27} \right)^{1/4}
  \frac{1}{c_0 \, \sqrt{s}}.
\end{align}
Expanding $N (r,s)$ from \eq{N3} for the $n=0$ branch in $r - r^*$
yields
\begin{align}
  N (r,s) \bigg|_{r - r^* \ll 1/\sqrt{s}, \ \ \ r > r^*} \, \approx \,
  1 - \exp \left\{ - \frac{\sqrt{\lambda}\,a \, 2^{1/4} \,
      s^{3/4}}{\pi\,c_0^{1/2} \, 3^{1/8}} \, \sqrt{r - r^*}\right\}.
\end{align}
(We have not expanded the exponent here and in previous expansions as
$a$ may be large.)  As $r^* \sim 1/\sqrt{s}$, we obtain $Q_s^2 \sim
1/r^{* \, 2} \sim s$, as observed in \eq{fit2}. The $Q_s^2 \sim s$
scaling is somewhat unusual, as in terms of $x_{Bj}$ it implies the
existence of saturation Bjorken-$x$, some $x^{\text{sat}}_{Bj}$, such
that the saturation sets in for $x_{Bj} \le x^{\text{sat}}_{Bj}$
independent of $Q^2$. However, as the $Q_s^2 \sim s$ scaling is
observed for large-$x_{Bj}$, we can not apply this conclusion for
small-$x_{Bj}$ physics, which we are studying using the Wilson loop.

One can easily check using Eqs. (\ref{asym}) and (\ref{N3}) that the
$s \, r^2 \gg 1$ asymptotics of \eq{N3} is given by \eq{Nlarge}. Thus
the high energy asymptotics of both possible solutions discussed here
is the same. The saturation scale at large $s$ is also the same for
$n=1$ branch and for the superposition of $n=2$ and $n=0$ branches. It
is given by \eq{qssmall}. Therefore, the main difference between the
two possible solutions is at small $r$/low energy $s$.

Let us finally note that, in DIS, for the Wilson loop to span the
whole shock wave in the longitudinal direction, the interaction of the
$q\bar q$ pair should be coherent over the longitudinal length of the
whole shock wave. In other words, in the rest frame of the nucleus,
the longitudinal coherence length of the $q\bar q$ pair should be
larger than the diameter of the nucleus. The standard condition for
this to happen in DIS is (see \cite{Kovchegov:2001dh} and references
therein)
\begin{align}
  \frac{1}{2 \, m_N \, x_{Bj}} \, \gg \, 2 \, R
\end{align}
with $m_N$ the nucleon mass.  Omitting factors of $2$, putting $m_N
\sim \Lambda$ and using $x_{Bj}$ from \eq{xbj1} yields
\begin{align}\label{coh_length}
  s \, r^2 \, \gg \, A^{1/3}.
\end{align}
Violation of this condition does not imply a breakdown of our
approximation. It would only mean that the Wilson loop does not
stretch across the whole shock wave in the longitudinal direction, but
only over a part of it. This would imply that, when the condition
(\ref{coh_length}) is violated, $A^{1/3}$ in our above expressions for
$N (r,s)$ should be replaced by $s \, r^2$. While this prescription
would modify some of our large-$x_{Bj}$ conclusions above, such as
\eq{Nsmall} along with the shape of the rise of the saturation scale
in \fig{qs1}, our main small-$x_{Bj}$ conclusion of constant $Q_s$ at
high energy (\ref{qssmall}) would remain the same.


\section{Conclusions}
\label{conc}

Let us restate our main results. We have studied DIS on a large
nucleus in ${\cal N} =4$ SYM using the AdS/CFT correspondence.  We
modeled the nucleus by an (infinitely) long shock wave. We argued that
since for DIS in QCD the incoming virtual photon splits into a
quark--anti-quark pair, which at high energies interacts with the
nucleus eikonally (without recoil), the DIS cross section can be
related to the expectation value of the Wilson loop, as shown in Eqs.
(\ref{sigN}), (\ref{N}) and (\ref{S}). We have calculated the
expectation value of the Wilson loop by minimizing the area of the
world-sheet for an open string connecting the quark and the anti-quark
lines in the loop. We obtain two physically possible solutions
resulting from different complex branches of string coordinates in the
bulk.  The two results for the forward scattering amplitude $N (r,s)$
are given in Eqs. (\ref{N1}) and (\ref{N3}), which are the main
results of the paper. They are plotted in Figs. \ref{ndip1} and
\ref{ndip2}. The corresponding saturation scales are plotted as a
function of energy in Figs. \ref{qs1} and \ref{qs2}. In both cases the
main feature is that the saturation scale becomes independent of
energy/Bjorken-$x$ at sufficiently high energies/small values of
$x_{Bj}$.

\FIGURE{\includegraphics[width=12cm]{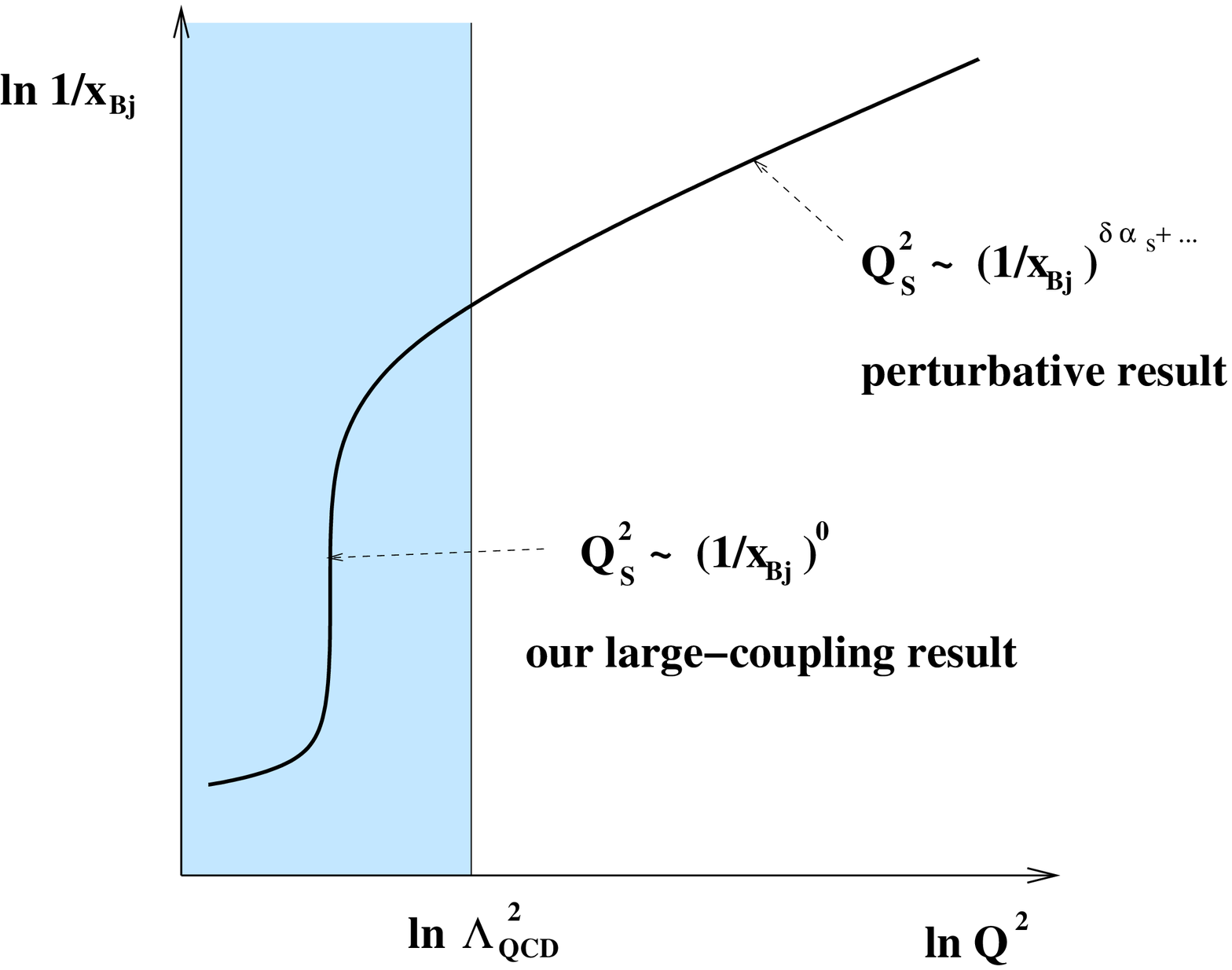}
  \caption{A sketch of the saturation line in the $(\ln Q^2, \ln 1/x_{Bj})$ plane. 
    In the small coupling regime ($Q^2 \gg \Lambda_\text{QCD}^2$) the
    saturation scale grows according to CGC predictions. In the large
    coupling regime ($Q^2 \lesssim \Lambda_\text{QCD}^2$) the
    saturation scale is a constant function of $x_{Bj}$ in accordance
    with our results in this work.}
  \label{satmap}
}

We interpret this result as follows. At small gauge coupling, say in
the BFKL evolution, each gluon emission is suppressed by a power of
$\as$, but is enhanced by a power of $\ln s$, which makes the
resummations parameter $\as \, \ln s$ of the order of $1$. High energy
$s$ is needed to generate this enhancement. However, at large coupling
the enhancement by logarithms (or by any other functions) of energy is
not needed anymore, as $\lambda \gg 1$. Hence one might conjecture
that at large $\lambda$ partons fill up the whole phase space just due
to sheer strength of the coupling. This way no energy enhancement is
needed and increasing the energy should not change anything in this
picture. This would lead to saturation scale being independent of
energy.

As our calculation was done for ${\cal N} =4$ SYM, it is important to
understand its implications for QCD. In an effort to do so we have
plotted the saturation line in the $(\ln Q^2, \ln 1/x_{Bj})$ plane in
\fig{satmap}. The traditional physics of CGC happens in the small
coupling regime of QCD for $Q^2 \gg \Lambda_\text{QCD}^2$, where
$\Lambda_\text{QCD}$ is the QCD confinement scale. The saturation
scale there grows with $1/x_{Bj}$ as some power $Q_s^2 \sim
(1/x_{Bj})^{\delta \, \as + \ldots}$ in the LLA, with ellipsis
denoting the higher order corrections. The strong coupling limit of
QCD is $Q^2 \lesssim \Lambda_\text{QCD}^2$. If one assumes that the
large-coupling analysis carried out in this paper is applicable to the
non-perturbative large-coupling $Q^2 \lesssim \Lambda_\text{QCD}^2$
sector of QCD, then we can conclude that in this region $Q_s^2 \sim
(1/x_{Bj})^0$. This is pictured in \fig{satmap}. In the
non-perturbative region we have simply sketched the curve from Figs.
\ref{qs1} and \ref{qs2}: saturation scale starts out growing with
decreasing $x_{Bj}$, but then levels off to a constant.  As $x_{Bj}$
decreases further, perturbative saturation mechanism would turn on and
small transverse size partons would be produced copiously due to $\ln
1/x_{Bj}$ enhancement, leading again to the saturation scale growing
as a positive power of $1/x_{Bj}$.  Indeed \fig{satmap} is applicable
only for a relatively small/dilute target, which would insure that
there is a region of small $x_{Bj}$ where $Q_s^2$ is still in the
non-perturbative region. For a very large nucleus or at very high
energies perturbative QCD effects would dominate making $Q_s^2$
perturbative. At the modern-day accelerators, such as HERA, Tevatron,
RHIC, and LHC, one gets $Q_s \approx 1 \div 3$~GeV for protons and
nuclei \cite{Albacete:2007sm}. This scale may not be large enough to
completely place us in the perturbative QCD region. Therefore it is
likely that the physics at these machines lies in the transition
region between the perturbative and non-perturbative regimes pictured
in \fig{satmap}.


\section*{Addendum}

After the paper was already in print, we realized that by modifying
certain mathematical conventions one could obtain the dipole
scattering amplitude $N (r)$ given in Fig. 9 from a {\sl single
  branch} of the solution for $z_{max}$ as opposed to the
superposition of two branches described in Sect. 4.3. A more
conventional definition of the Nambu-Goto action in Minkowski space is
different by a minus sign from our Eq. (2.7). Using only the $n=2$
root for $z_{max}$, denoted $z_{max}^{n=2}$, in such action along with
Eq. (4.29) would lead to the dipole amplitude in Fig. 9.
Alternatively, using the complex conjugate of the $n=2$ root for
$z_{max}$, denoted $z_{max}^{n=2 \, *}$, in Eq.  (4.30) would also
give the dipole amplitude in Fig. 9 coming from a single branch of the
solution. This observation does not modify any of the results in the
paper (and in particular in Sect. 4.3).


\acknowledgments

We would like to thank Hong Liu, John McGreevy, Krishna Rajagopal, and
Amit Sever for interesting discussions and for encouraging us to write
this paper. We are also very grateful to Samir Mathur and Al Mueller
for a number of very informative and supportive discussions. We thank
Al Mueller for a careful reading of the manuscript.

This research is sponsored in part by the U.S. Department of Energy
under Grant No. DE-FG02-05ER41377.


\appendix

\renewcommand{\theequation}{A\arabic{equation}}
  \setcounter{equation}{0}
\section{Monotonicity of $N(r,s)$ as a function of $r$}
\label{A}

We wish to show that $N(r, s)$ appearing in Fig. \ref{ndip1} is a
monotonic function of $r$ for all $s$, and not just for the values of
$s$ used in the plot. In particular, we claim that $N(r, s)$ goes to
zero when $r \rightarrow 0$ and $N(r, s)$ goes to $1$ when $r
\rightarrow \infty$ with a monotonic interpolation between the two
extremes and without any local minima or maxima in-between. This would
imply that $N(r, s)$ is not only monotonic, but in addition is
restricted between zero and one, a necessary condition arising from
unitarity.

The minimum requirement we need to prove is that the derivative of the
expression in the square bracket in the exponential of (\ref{N1}) is
positive for any (positive) values of $r$ and $s$. We should have in
mind that $\rho_{max}=\rho_{max}(r,s)$ [see Eqs.  (\ref{rho}),
(\ref{param}-\ref{delta})].  We denote by $E(r, s)$ the expression in
the square brackets in the exponent of (\ref{N1})
\begin{align}\label{E}
  E(r,s) \, \equiv \, \frac{c_0^2\,r^2}{\rho_{max}^3(r,s)} +
  \frac{2}{\rho_{max}(r,s)} - \frac{2}{z_{h}(s)}.
\end{align}
We proceed by evaluating its derivative with respect to $r$
\begin{align}\label{dE}
  \frac{\partial E}{\partial r} \,= \, - \left(
    \frac{3\, c_0^2 \,r^2}{\rho_{max}^4} + \frac{2}{\rho_{max}^2} \right)
  \frac{\partial \rho_{max} }{\partial
    r}+\frac{2\, c_{0}^2 \, r}{\rho_{max}^3}
\end{align}
At this point we recall that $z_{max} = i \rho_{max}$. \eq{zm}
then gives 
\begin{align}\label{rhoeq}
  c_0^2 \, r^2 \, = \, s^2 \, \rho_{max}^6 -  \rho_{max}^2.
\end{align}
Differentiating both sides of \eq{rhoeq} with respect to $r$ we obtain
\begin{align}\label{dzm}
  c_0^2 \, r \, = \, \rho_{max} \, \left( 3 \, s^2 \, \rho_{max}^4 - 1
  \right) \, \frac{\partial \rho_{max} }{\partial r}.
\end{align}
Solving \eq{dzm} for $\frac{\partial \rho_{max} }{\partial r}$ and
eliminating $\frac{\partial \rho_{max} }{\partial r}$ in \eq{dE}
yields
\begin{align}\label{dE1}
  \frac{\rho_{max}^3}{c_0^2 r} \frac{\partial E}{\partial r} \, = \, -
  \frac{2 \, \rho_{max}^2+3 \, c_0^2 \, r^2} {3 \, s^2 \,
    \rho_{max}^6-\rho_{max}^2}+2.
\end{align}
We now eliminate $\rho_{max}^6$ in the denominator of \eq{dE1} using
\eq{rhoeq}. This yields
\begin{align}\label{dE2}
  \frac{\rho_{max}^3}{c_0^2 \, r} \, \frac{\partial E}{\partial r} =
  -\frac{2\rho_{max}^2 + 3 \, c_0^2 \, r^2} {3 \, (c_0^2
    r^2+\rho_{max}^2)-\rho_{max}^2}+2 \, = \, 1 \, > \, 0.
\end{align}
This completes the proof of our statement. As can be easily shown (see
also Fig. \ref{reimr}) $\rho_{max} > 0$. Therefore we conclude that
\begin{align}
   \frac{\partial E}{\partial r} > 0
\end{align}
for all $r >0$, which means that $N (r,s)$ in \eq{N1} is a
monotonically increasing function of $r$.



\providecommand{\href}[2]{#2}\begingroup\raggedright\endgroup


\end{document}